\documentclass[acmsmall]{acmart}
\AtBeginDocument{%
  }

\setcopyright{rightsretained}
\acmJournal{THRI}
\acmYear{2026} \acmVolume{1} \acmNumber{1} \acmArticle{}
\acmMonth{1} \acmPrice{} \acmDOI{10.1145/3796524}

\acmJournal{JACM}
\acmVolume{37}
\acmNumber{4}
\acmArticle{111}
\acmMonth{8}

\usepackage{array}
\usepackage{longtable}
\newcolumntype{x}[1]{>{\centering\arraybackslash\hspace{0pt}}p{#1}}

\begin{document}

\title{Eye-Tracking-Driven Shared Control for Robotic Arms: Wizard of Oz Studies to Assess Design Choices}

\author{Anke Fischer-Janzen}
\email{anke.fischer-janzen@hs-offenburg.de}
\orcid{0000-0003-3992-3163}
\affiliation{%
  \institution{Offenburg University of Applied Sciences}
  \city{Offenburg}
  \state{Baden-Württemberg}
  \country{Germany}
}

\author{Thomas M. Wendt}
\affiliation{%
  \institution{Offenburg University of Applied Sciences}
  \city{Offenburg}
  \country{Germany}
}
\email{thomas.wendt@hs-offenburg.de}

\author{Daniel Görlich}
\affiliation{%
  \institution{Offenburg University of Applied Sciences}
  \city{Offenburg}
  \country{Germany}
}
\email{daniel.goerlich@hs-offenburg.de}

\author{Kristof Van Laerhoven}
\affiliation{%
  \institution{University of Siegen}
  \city{Siegen}
  \country{Germany}}
\email{kvl@eti.uni-siegen.de}

\renewcommand{\shortauthors}{Fischer-Janzen et al.}

\begin{abstract}
    Eye-tracking-driven controls for assistive robotic arms provide people with severe physical disabilities with intuitive interaction opportunities.In this context, shared control can improve acceptance of the robot by partially automating task execution. Based on recent literature, we present a Wizard of Oz design for shared control driven by eye-tracking. This approach allows for the rapid exploration of user expectations to inform future design iterations. Two studies were conducted to assess user experience, identify design challenges, and determine ways to improve usability and accessibility. The first study involved people with severe motor disabilities (PSMD) and consisted of an online survey. The second study aimed to gain technical insights through direct robot interaction to provide a comprehensive overview of the findings.
\end{abstract}



\received{20 February 2007}
\received[revised]{12 March 2009}
\received[accepted]{5 June 2009}

\maketitle 

\section{Introduction}
\label{sec:Introduction}

People with severe motor disabilities (PSMD) often rely on hands-free control modalities, such as Brain-Computer Interfaces, sip-and-puff systems, and eye-tracking interfaces. These technologies have been successfully used by individuals with conditions such as locked-in syndrome, cerebral palsy, late-stage amyotrophic lateral sclerosis, or multiple sclerosis. As control inputs for assistive robotic arms, they can support independence, self-determination, and privacy in daily life, reducing reliance on family members and caregivers \cite{FischerJanzen2024, Awuha2024, Baumeister2021}. 

A shared control allows users to give the robot high-level commands, such as selecting an object or goal, by automating task sequences. In eye-tracking-driven approaches, advances in AI allow us to move beyond manual mappings toward intelligent interfaces that interpret gaze patterns to extract task intent \cite{Aronson2018, Belardinelli2024, Gonzalez2024}. Environmental context, such as the presence of objects, can further improve autonomous decision-making \cite{Zhang2025, Yang2021}. These advances have led to the development of various robot control methods for Human-Robot Interaction (HRI) that differ in terms of control strategies \cite{Mohebbi2020, FischerJanzen2024}. With a higher level of robot autonomy, challenges such as controlling three-dimensional robot movement with two-dimensional eye movements can be resolved. Bypassing the need for mode switches, as required for direct control, to cover the 3-dimensional robot space, or independent joint control, can reduce workload and decrease the concentration required for the task \cite{FischerJanzen2024, Kronhardt2023, Herlant2016}. Shared control approaches have been shown to positively impact interaction perception and reduce task completion time \cite{Bhattacharjee2020, Kronhardt2023}. A challenge associated with a high level of automation is maintaining the sense of agency. Community feedback is essential to improve design and evaluate usability and accessibility \cite{Stalljann2020, Baumeister2021, Nanavati2024}, as well as to uncover influences and design controls with an appropriate level of autonomy. It provides firsthand insights into user expectations, thus enhancing acceptance rates \cite{Nanavati2024, Canal2021}. 

The novelty of this work is characterized by a detailed look at usability and functionality regarding robot acceptance parameters in respect to articulated robots used to assist in daily life. Current task intent recognition methods are based on object labeling done by object detection or segmentation, which we evaluated in this context on feasibility. This work incorporates community feedback on an initial shared control design to identify challenges and guide improvements. A Wizard of Oz design of the eye-tracking-driven control was developed and used in two studies. The first study was an online survey distributed to PSMD, their families, friends, and medical professionals. The goal of the study was to assess everyday usability and identify task preferences. The stakeholder group was defined as PSMD who experience significant motor impairments in their arms and torso. Previous research has shown differences in the perception of virtual versus real robot interactions \cite{Petrich2022}. A second study was conducted, in which participants without motor impairments and with varying levels of robotics experience interacted with the physical system. This study focused on deriving technical characteristics that can inform the constraints of a future control algorithm, such as feasible dwell times and accuracy requirements. The main contributions of this work are as follows: 

\begin{enumerate} 
    \item Presentation of an initial eye-tracking driven shared control design leveraging object detection algorithms for daily tasks.
    \item A functionality and usability evaluation, incorporating feedback from potential users, their families, medical professionals, and non-disabled participants with and without prior robot experience.
    \item Refinements and task preferences for the initial shared control design, derived from insights obtained in both studies.
\end{enumerate}

\section{Background}
\label{sec:Background}

\subsection{Gaze as control input in shared controls} 
\label{subsec:BackgroundSharedControl}
"Shared control in assistive robotics blends human autonomy with computer assistance, thus simplifying complex tasks for individuals with physical impairments" \cite{Goldau2024}. Such a system can achieve lower performance times and error rates when the user provides strategic intent and the autonomy supplies continuous, low-level stabilization, obstacle avoidance, and motion refinement \cite{Farhadi2025}. Advances in sensing, inference, modeling, and learning methods have led to novel developments in this research area \cite{Selvaggio2021}. 

Gaze can be used as an implicit or explicit control input for shared controls \cite{Loke2025}. As an implicit control input, natural gaze patterns are used to recognize or predict user intent. For example, gaze behaviors such as fixating on task-relevant objects before naming or manipulating them can be used and predicted before the desired task even begins \cite{Admoni2017, Aronson2020,  Belardinelli2024, Gonzalez2024, Selvaggio2021}. Huang and Mutlu demonstrated an example of such controls, in which ingredients were predicted by gaze patterns \cite{Huang2016}. Admoni et al. created a shared control system based on a POMDP (Partially Observable Markov Decision Process), which predicts single tasks from gaze patterns in an unstructured environment \cite{Admoni2016}. Zhang et al. presented another approach based on the usage of a large language model (LLM) and vision foundation model (VFM), that recognizes user intent by interpreting labels created by the VFM and using the LLM to find common object combinations \cite{Zhang2025}. Wang used machine learning to recognize actions based on 3D gaze-related features to control a robot \cite{Wang2021}. Lastly, Farhadi et al. presented a review of BCI-related controls and shared control approaches \cite{Farhadi2025}. Challenges in predicting patterns of natural gaze can arise from multiple cognitive processes occurring in our brain at any time. As Belardinelli states, this issue worsens when performing activities more complex than grasping or selecting an object \cite{Belardinelli2024}. If incorrect or imprecise predictions are made, collisions or large penalties in completion time and user preferences can occur \cite{Jain2019}. As Cao et al. point out, timely and accurate estimation of intentions and, consequently, prediction remain major challenges \cite{Cao2025}.      

Gaze, as an explicit input, can vary in its function in control. The first eye-tracking controllers for robotic arms were developed as early as 2001 \cite{Kim2001}. Since then, several systems have been developed, but only a few have become commercially available \cite{FischerJanzen2024, Suryadarma2024, Brose2010}. One example of explicit control input is directional control, in which the robot moves in the direction of gaze \cite{Alsharif2017, Sunny2021}. Graphical User Interfaces (GUIs) are available in which the robot is moved with gaze-selected buttons, taking advantage of more accurate stationary eye-tracking systems \cite{Dragomir2021, Stalljann2020}. Such controls are easy to learn, provide a good sense of agency, and reduce cognitive load \cite{Dragomir2021, Dziemian2016}. However, controlling each of the robot movements separately is more time consuming. A popular approach, as outlined in Belardinelli's work, is to use relevant objects in the scene and define them as Areas-Of-Interest (AOI). Fixations within these areas can be evaluated and, together with a certain dwell time, used as a selection method \cite{Catalan2017, Zeng2017, Tostado2016}.
We refer to the work of Fischer-Janzen et al. and Suryadarma et al. for examples in realization, due to the number of related works \cite{FischerJanzen2024, Suryadarma2024}. A major limitation of these works is that they do not allow for the subsequent selection of multiple tasks, which is a major requirement for daily use, as demonstrated by Hagengruber et al. \cite{Hagengruber2025}. In this work, gaze is used as explicit input by focusing on the desired object.

Challenges can occur in both implicit and explicit control inputs that need to be considered:
\begin{itemize}
    	\item \textbf{Unintended selection:} When the gaze is used as a pointing device, e.g., to select an object, rules must be applied to ensure that not all fixations do not trigger selections. These unintentional selections are described by the Midas Touch problem \cite{Duenser2015, Jones2018, Stalljann2020}. Reasons for this problem include unintentional and reflexive eye movements, overlapping objects, distraction, stress, and the limited accuracy of the eye-tracking device \cite{Holmqvist2017}. When objects are close together and the eye tracker's accuracy is insufficient, the gaze may be tracked to the wrong object \cite{Jones2018}.
    	\item \textbf{Context misinterpretation:} The aforementioned parallel cognitive processes can result in task ambiguity. In an interview conducted by Herr et al. \cite{Herr2016}, participants stated: “Well, now I look at you [the interviewer], but I actually focus on the bottle behind you. That would be difficult for the robot to recognize”. This behavior suggests that attention to a particular location is sometimes unrelated to intention or cognitive process.  
    	\item \textbf{Physiological factors:} There are some physiological factors that affect the accuracy of the gaze. Neurological conditions, fatigue, drooping eyelids, and strabismus can affect the functionality of eye-tracking devices \cite{Holmqvist2017}. 
\end{itemize}
Such challenges can lead to frustration and a high cognitive load for users trying to avoid them \cite{Duenser2015, Jones2018}.

\subsection{Characteristics in gaze-based shared controls}
\label{subsec:BackgroundCharacteristics}
Our system design was inspired by the impact factors on user acceptance of shared control found in the current literature. Many characteristics and technology acceptance models exist \cite{Beer2011}. Due to the multifaceted nature of user acceptance, we present only the factors that influenced our design. For more information on characteristics influencing user acceptance, see the works of Smarr et al., Rahman et al., Riek et al., and Beer et al. \cite{Smarr2012, Rahman2025, Riek2014, Beer2011}. We introduce three categories to facilitate an overview of the results: Technical challenges, level of autonomy, and user preferences. 

\textit{Technical challenges} include ensuring the system's reliability and functionality. The system must be robust in changing or unorganized environments, especially when the robot is mounted on an electric wheelchair \cite{Selvaggio2021, Stanger1994}. The system's capabilities related to task performance also define its functionality in everyday life. We discussed such challenges in Section \ref{subsec:BackgroundSharedControl}. The subtasks necessary to correctly complete a task must be understood and set in the proper context relative to each other. 

The \textit{level of autonomy} is a subject of ongoing debate and can range from tele-operated systems to fully autonomous robots \cite{Beer2014}. Recent studies have shown that more autonomy does not necessarily align with user preferences \cite{Nanavati2024b, Bhattacharjee2020}. If a robot with a high level of autonomy produces more errors in task execution, users tend to reject the system \cite{Bhattacharjee2020}. This can lead to a loss of trust or a reduced sense of agency due to the system's unpredictable behavior \cite{Nanavati2024b, Baptista2025}. Losey et al. summarized this as a duality: More automation reduces the user's burden when performing tasks, but it also limits the mutual support when it ignores the user’s intent. 
The autonomous behavior reduces the operator's workload when performing repetitive tasks \cite{Selvaggio2021}. As Loke et al. state, high mental workload contributes to the abandonment of assistive devices \cite{Loke2025}. Due to the potential shift in cognitive resources toward the task and away from precise robot movements, control can benefit from being easy to understand \cite{Mazzola2025, Selvaggio2021}. However, this may vary depending on the complexity of the task \cite{Stanger1994, Shafti2019, Goldau2024}, leading back to the initial question of what level of autonomy is most efficient and reliable in assisting the user.

\textit{Users' preferences} vary based on personal metrics such as the user's age, level of disability, living arrangements, and familiarity with technology \cite{Stanger1994, Smarr2012, Nanavati2024b}. Incorporating user preferences influences perception of and satisfaction with the system \cite{Cakmak2011, Canal2021}. Parameters such as speed can influence the user perception \cite{Bhattacharjee2020}. Another example is the participant's preference for handing over an object with a particular orientation \cite{Cakmak2011}. The ability to adapt the system seamlessly to users' preferences regarding the level of assistance has given rise to shared autonomy approaches. Those are based on the autonomy's understanding of the user's intentions and of the surrounding environment \cite{Selvaggio2021}. In this context, research focuses on arbitration strategies that adapt the level of autonomy to the user's preferences, as demonstrated in the works of Dragan et al., Losey et al., and Jain et al. \cite{Dragan2013,Jain2019, Losey2018}. The arbitration scheme, together with the interaction paradigm, describes the collaboration between human and robot \cite{Farhadi2025}.

When designing well-balanced shared control, a trade-off must be found for all technical and ethical factors. Section \ref{subsec:MethodsSCDesignChoices} presents a solution that considers and balances the different influences.

\subsection{Tasks in assistive robot manipulation}
\label{subsec:BackgroundTaskPreference}
Edemekong et al. describe Activities of Daily Living [(ADLs)] as basic routine tasks that most healthy individuals can perform without assistance \cite{Edemekong2025}. ADLs are often used to evaluate a person's level of disability, as in the WHODAS 2.0 \cite{WHO2010b}. Nanavati et al. used ADLs in their review to compare the responses of individuals requiring ADL assistance with the work of researchers in the field of assistive robotics. A major misalignment was found. While most researchers focus on mobility and eating, people in need of assistance state that they need help with going out, managing money, doing housework, preparing meals, bathing, and dressing \cite{Nanavati2024b}. Petrich et al. took a different approach, examining the behavior of people in everyday life and uncovering results quite different from those previously mentioned. In their study, tasks such as reaching to pick up and place items, opening and closing doors, and using switches and buttons were most important, according to video analysis. These results differed from the participants' survey answers \cite{Petrich2022}. These results highlight that different approaches can reveal various aspects of task prioritization and preferences in HRI. Smarr et al. explored robot acceptance among the elderly and found that the elderly accept robots more readily when they assist with learning new skills and hobbies. However, human support was preferred for communication, which emphasizes the importance of social contact \cite{Smarr2012}.
The examples presented above show that assistive robotic arms cannot directly assist with activities such as bathing or managing finances. It is important to distinguish between robotic arm-related tasks and others. 

As stated by Hagengruber et al., the purpose of assistive robotics is to carry out various tasks in succession in order to depict a realistic daily routine \cite{Hagengruber2025}. They presented an sEMG-based control system that could perform a drinking task, including retrieving a cup from a drawer and pouring a drink. Chi et al. demonstrated that dividing complex tasks into subtasks facilitates accomplishment. For example, they divided an eating task into subtasks such as getting a spoon, scooping food, and bringing it near the user’s mouth \cite{Chi2025}. Using gaze as a control input, one can obtain gaze patterns specific to each subtask that can be used to train AI models \cite{Hayhoe2014}.

Robots can perform daily tasks, such as household chores and personal hygiene, as well as medical procedures \cite{Herr2016, Rahman2025, Bien2003, Rahman2025, Wang2020}. The preferred tasks vary depending on different influences, such as the user’s attitude towards a task. Delegating household tasks to robots is one example of this \cite{Herr2016}. Similarly, Petrich et al. suggested automating tasks that are not prioritized by the community to reduce cognitive load \cite{Petrich2022}.

Lastly, one must determine whether a robotic arm is the right tool for the task at hand. For example, Petrich et al. stated that robots can help with using electronics, such as smartphones; however, it is more convenient to use a hands-free phone \cite{Petrich2022}. Similar results were found for toileting, where people preferred the aid of caregivers and smart home applications. Hagengruber et al.'s work, in particular, demonstrates the advantage of using multiple assistive devices together. For example, it simplified opening a drawer by holding the handle with the robot and moving the wheelchair backward \cite{Hagengruber2025}.

In addition to the work of Nanavati et al. \cite{Nanavati2024b}, our review revealed that the four most frequently mentioned tasks, suitable for assistive robots, were food related tasks (serving a meal, eating) \cite{Bien2003,Stanger1994, Jardon2012, Chang2003, Canal2021, Bhattacharjee2020, Nanavati2024}, drinking related tasks \cite{Bien2003,Stanger1994, Jardon2012, Chang2003}, picking up objects \cite{Bien2003,Stanger1994,Jardon2012, Chang2003, Nanavati2024b}, and personal hygiene \cite{Stanger1994, Jardon2012, Nanavati2024b}. Personal hygiene included various tasks such as shaving \cite{Bien2003, Chang2003}, washing the face and hands, brushing teeth, combing hair, shaving, applying makeup \cite{Jardon2012}, and dressing \cite{Canal2021, Nanavati2024b}. The importance of being able to manipulate a wide variety of objects benefits the ability to get around, e.g., with an electric wheelchair, by being able to move blocking objects out of the way \cite{Nanavati2024b}.

Other miscellaneous tasks were mentioned with taking medicine \cite{Nanavati2024b}, working \cite{Nanavati2024b}, leisure activities \cite{Stanger1994, Nanavati2024}, turning on and off switches \cite{Bien2003, Chang2003}, opening and closing doors \cite{Bien2003, Brose2010, Chang2003}, scratching oneself \cite{Bien2003, Chang2003}, and making tea. Sources from 2003 show that changing CDs and removing paper from fax machines are examples of how task preferences change over time \cite{Bien2003, Chang2003}. 
As the accessibility of smart home applications continues to evolve, it is expected that more tasks will become less important for robot assistance, such as flipping light switches or opening doors. 

Based on these results, tasks were selected for the Wizard of Oz design. In Section \ref{subsec:ResultsOnlineSurvey}, the literature findings were used to group the responses from the online survey. In Section \ref{subsec:DiscussionUsability}, they were used to compare the results in terms of changes in task prioritization over time.

\subsection{Participation in assistive robot design}
\label{subsec:BackgroundParticipation}

The previous section presented characteristics that influence user acceptance and robot abandonment. Several authors argue that involving the community and people with disabilities in the design of assistive devices is more advantageous than involving people without disabilities. Including PSMD in the design process of assistive devices helps designers gain insight into the lives and needs \cite{Baumeister2021, Stanger1994, Nanavati2024}. Arboleda et al. point out that end users' perspectives can differ from those of other groups, such as caregivers \cite{Arboleda2021}. As Stalljann et al. demonstrate, control inputs are experienced differently by disabled and non-disabled participants \cite{Stalljann2020}. Most studies on eye-tracking controls recruit non-disabled people as participants \cite{FischerJanzen2024}. Reasons PSMD are rarely included in studies include recruitment challenges, small sample sizes, and transportation logistics \cite{Nanavati2024}. 

Nanavati et al. \cite{Nanavati2024} discuss approaches based on three-dimensional key features to determine suitable study designs for achieving the study's objective.  The first dimension determines whether participation is at the individual or community level. Individual participation provides the opportunity for deeper insight into the user’s life and preferences, while community work provides insight from many participants. This dimension also affects the time required to conduct the study and the connection between participants and experimenters. The second dimension highlights the logistical effort required of participants and researchers. PSMD participants may require assistance with transportation to study sites, which can facilitate conducting studies at the participants' locations. Arboleda et al. stated that conducting the study itself might disrupt participants' daily routines, which could induce stress \cite{Arboleda2021}. Therefore, other possibilities can be considered, such as online interviews. Depending on the robot setup, it may be easier to conduct studies at the participant's location to reduce the need for special transportation and availability of medical professionals. The third dimension describes the benefits to researchers and the community. PSMD, their families, and medical professionals are intrinsically motivated to participate in the design process of assistive devices and robots. They can benefit and contribute by providing insight to adapt the systems to their needs.

These findings demonstrate the value of involving PSMD in the design process and influenced the study design in Section \ref{subsec:MethodOnlineSurvey}.

\section{Methods}
In Sections \ref{sec:Introduction} and \ref{sec:Background}, we presented the influencing factors and user challenges that affect acceptance of eye-tracking-driven shared controls. These factors and challenges were incorporated into the system design presented in Section \ref{subsec:MethodsSCDesignChoices}. 
Several groups of interest were interviewed with two study designs, to obtain broader feedback. The setup described in Section \ref{subsec:MethodsWOz} was used in both studies. An online survey was sent to stakeholder groups, their families and medical professionals. The results are presented in Section \ref{subsec:MethodOnlineSurvey}. Measurements were conducted to verify technical functionality and participants' perception of the system interacting directly with the robot (see Section \ref{subsec:MethodHandsOnStudy}).

\subsection{Shared control design choices}
\label{subsec:MethodsSCDesignChoices}

The initial idea for the system is to include a shared control, which automates parts of the robot movement, such as trajectory planning, grasping, and object manipulation. The level of autonomy is presented according to the guidelines of Beer et al. (2014) \cite{Beer2014}. The following considerations were used to create the Wizard of Oz design and describe the interaction to the participants.

\textbf{What task is the robot to perform?}
This shared control system is designed to assist PSMD with daily tasks. Ongoing work is developing a framework that can be expanded to include new ADL, thereby increasing the range of tasks it can perform. The first approach includes the tasks of picking up food with a fork, turning on a light switch, picking up and placing a block, scooping from a bowl, pouring water into a glass, and handing it to the participant. These tasks have to be performable in unstructured environments, such as at home and in public places like restaurants. 

As shown in Section \ref{subsec:BackgroundTaskPreference}, these tasks are preferred for robotic assistance in everyday life. 

\textbf{What aspects of the task should the robot perform?}
The chosen input modality is eye-tracking, because eye movement is often unrestricted or only slightly restricted in cases of locked-in syndrome or cerebral palsy, in contrast to other motor functions, such as arm movement. The robot is controlled by fixating the desired object, and the robot decides which available task to perform based on the selected object. Selecting multiple objects with the gaze makes task selection more robust. For instance, looking at a cup could indicate that the user wants to drink from it, move it, fill it, or combine available tasks. Adding the ability to combine the cup with other objects, such as a bottle or a location, distinguishes the task as refilling the cup or placing it in the selected location. Based on the information about the object and the task, the robot performs the task autonomously. The robot determines the object's location, selects a suitable grasp related to the task, and calculates the optimal trajectory. 

\textbf{To what extent can the robot perform those aspects of the task?}
The user is always in control of the robot in order to maintain a sense of agency (discussed in \cite{Nanavati2024b}). Validation methods are planned to intervene in case of incorrect task execution before the robot begins (suggested in \cite{Baptista2025}). Settings are planned to reduce the robot's velocity, if desired (suggested by \cite{Bhattacharjee2020}). Implementing dwell time for selecting objects has the potential to reduce incorrect interactions with objects through natural gaze behavior, thus minimizing the Midas Touch problem.

\textbf{What level can the robot's autonomy be categorized?}
According to the classification of Beer et al. \cite{Beer2014}, robot autonomy is categorized as Shared Control with Human Initiative. 

\textbf{How might autonomy influence HRI variables?}
The system must be reliable and functional. User safety is part of ongoing discussions and must be ensured. The system is designed to be easy to learn and use. In real-world settings, objects are selected by focusing on them. However, the accuracy of the eye tracker limits the correct selection of the desired object when gaze is tracked outside the area of interest (AoI). This raises the question of how accuracy errors influence the functionality of this approach. This will be tested using the Wizard of Oz setup.

\subsection{Wizard of Oz design}
\label{subsec:MethodsWOz}
\begin{figure}[ht]
  \centering
  \includegraphics[width=0.9\linewidth]{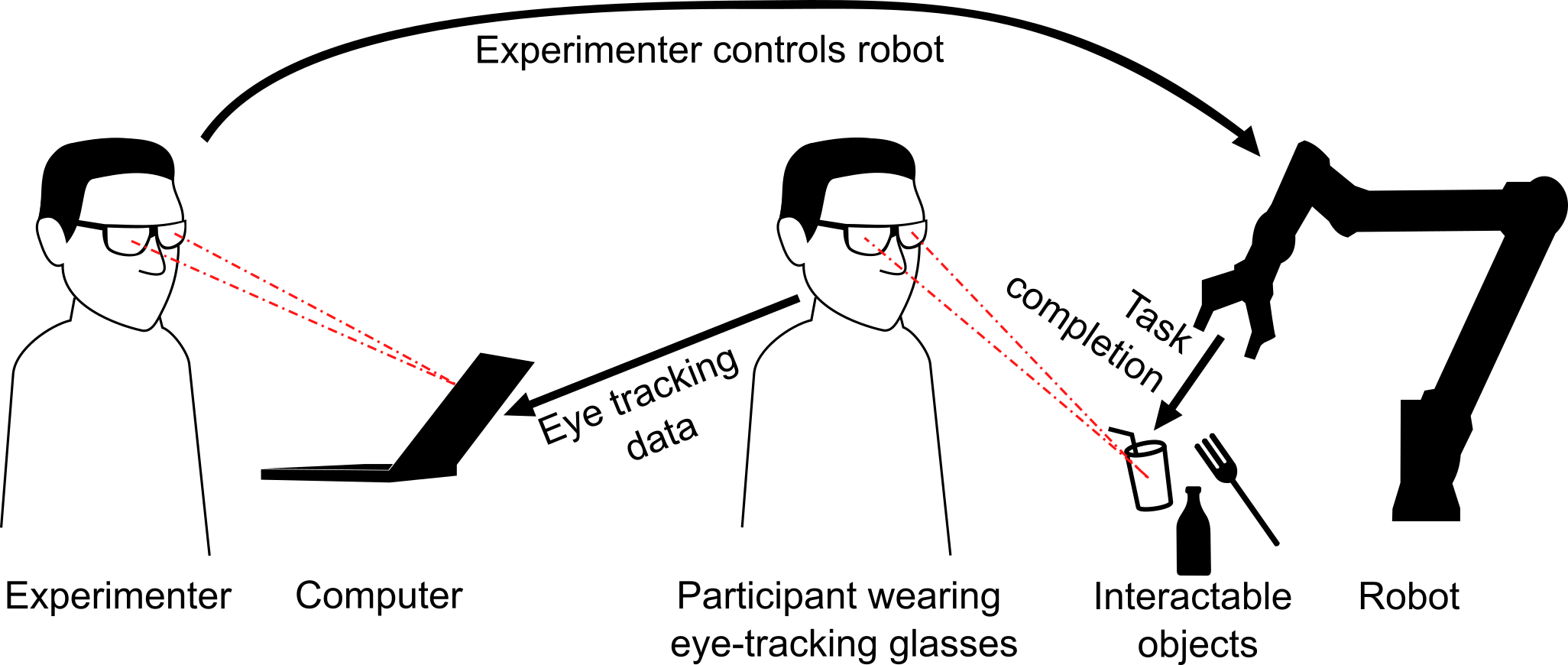} 
  \caption{Wizard of Oz design: The participant selects a task by focusing his or her gaze on the desired object. The gaze is visualized on the computer by displaying the live-video of the eye-tracker. The experimenter starts the task via the robot's web interface. The robot then performs the task and returns to the home position to wait for further commands. The home position is used as indication that the participant can select the next task.} 
  \Description{Wizard of Oz design: The participant selects an object by focusing on it with his or her gaze. Gaze is visualized on the computer by showing the eye-tracker's live-video and including a red circle where the user currently looks. Without any further verbal communication the experimenter starts the task via the robot's web interface by supervising the gaze movement of the participant. Afterwards, the robot performs the task and returns in the start position to wait for further commands. The home position is the indication that the participant can select the next task.} 
  \label{fig:figure1}
\end{figure}

\begin{figure}[ht]
  \centering
  \includegraphics[width=1.0\linewidth]{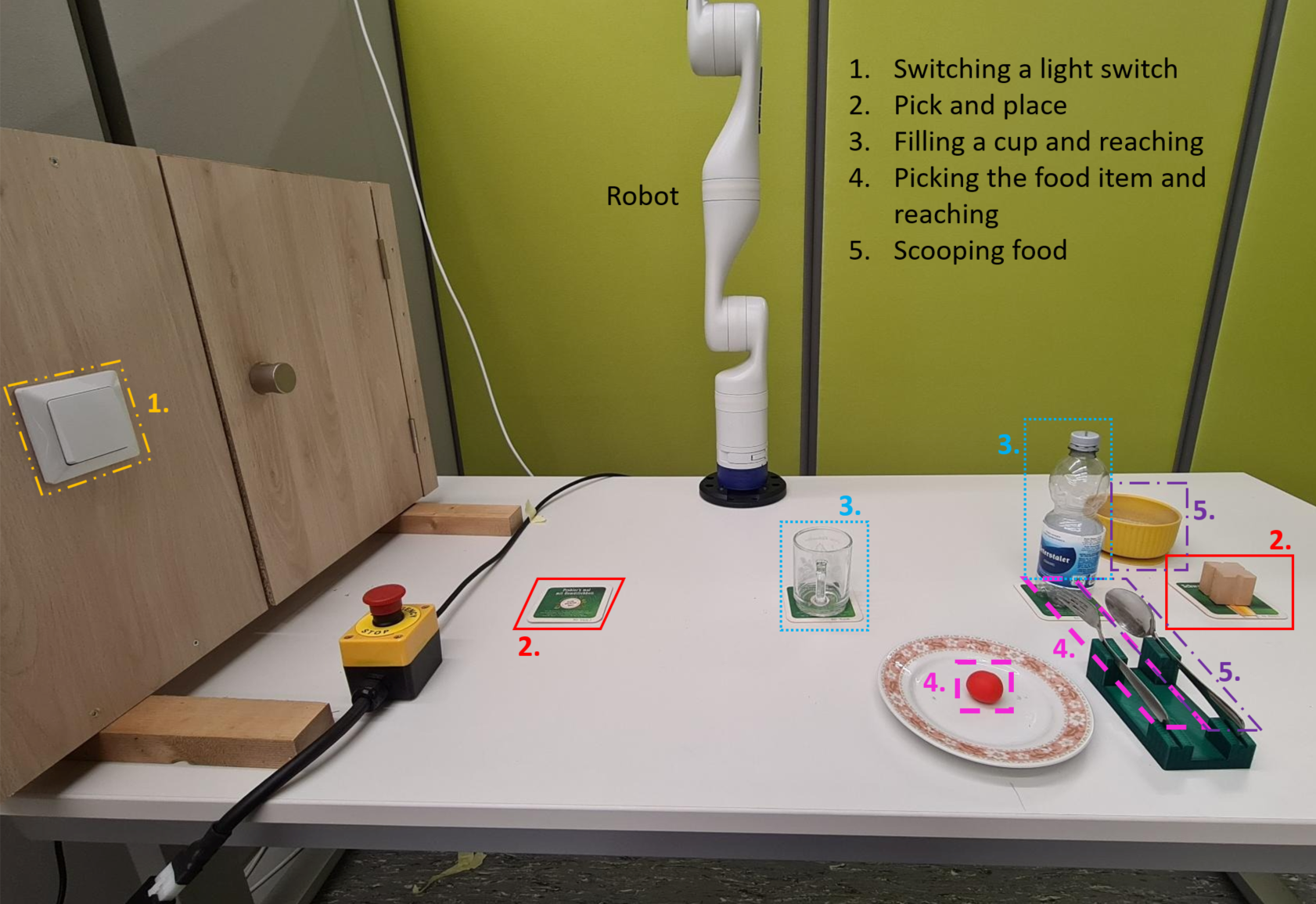}
  \caption{Nine interactable objects were presented with which five tasks were performed. Participants were shielded from visual distractions by green walls surrounding the experimental area.} 
  \Description{The figure shows the setup with nine objects placed either on the table or, in the case of the light switch, on a plywood wall that could be selected to perform five tasks. The scene is seen from the participant's point of view. The robot was mounted on the opposite side of the table, slightly off-center to the left of the participant. The first task described switching a light switch. The plywood wall with the light switch was on the far left side of the table. The second task, Pick and Place, could be started by selecting either the platform between the light switch and the robot or the block on the far right side of the table. Filling a cup and reaching it to the participant described the third task and was selected by looking at either the cup or the bottle placed next to each other at the same depth of the platform, slightly to the right of the robot. The fourth task was initiated by selecting either the food item placed on a plate in front of and between the cup and bottle, or the fork right next to it. Lastly, scooping food with a spoon was selected by looking at the spoon to the right of the fork or a bowl placed behind the block. Participants were shielded from visual distractions by green walls surrounding the experimental area. The emergency stop button was placed on the table in the front left corner in front of the light switch so that participants could reach it if they felt unsafe.} 
  \label{fig:figure2}
\end{figure}

An early system evaluation was conducted to determine whether the design meets the community's expectations for the found influencing factors. This approach shortens development time by intervening early in the event of low acceptance or functional challenges. In the Wizard of Oz design, the experimenter performs the autonomously controlled robot tasks, as seen in Figure \ref{fig:figure1}. Table \ref{tab:tab2} presents the Wizard of Oz system according to the guidelines presented by Riek \cite{Riek2012}. Figure \ref{fig:figure2} shows the five selectable tasks with the Wizard of Oz system. A video of task performance was created and implemented in the online survey to demonstrate the system's functionality (Section \ref{subsec:MethodOnlineSurvey}). This video is included in the Supplementary Materials. The system was also used in the hands-on study, where participants operated the system in a controlled environment (Section \ref{subsec:MethodHandsOnStudy}). Both studies were reviewed and approved by the ethics committee of the first author's institution.

\begin{table} [ht]
  \caption{Wizard of Oz presentation after Riek \cite{Riek2012} stating experimental components to the robot, user, wizard and to general information.}
  \label{tab:tab2}
  \begin{tabular}{cx{4.5cm}x{4.5cm}}
    \toprule
    Experimental Component &  Online survey & Hands-on study\\
    \midrule
    Robot      &  \multicolumn{2}{c}{7-DoF Kinova Gen3}\\
    Level of autonomy   & \multicolumn{2}{c}{Teleoperation} \\
    Robot capabilities  & Food item, scooping food, filling cup & Food item, pick and place, filling cup, light switch\\
    \midrule
    No. of participants        & 34 & 24 \\
    Demographics       & Age: $40\pm 27$;       & Age: $30.5\pm 8$;   \\
    & 14 PSMD, 9 family members, 11\,medical professionals; & 11 non-prior experience group, 13 prior-experience group;\\
     & 14 male, 17 female, 3 diverse & 7 female,17 male\\
    Instructions   & See Section \ref{subsubsec:MethodsOnlineSurveyInstructions} & See Section \ref{subsec:MethodsHandsOnInstructions} \\   
    Convincing simulation   & Not asked & Yes \\  
    \midrule
    Wizard   & \multicolumn{2}{c}{Experimenter took the role of the wizard}  \\  
    Familiar with hypothesis?   & \multicolumn{2}{c}{Yes} \\ 
    Production variables  & - & See Section \ref{subsec:MethodsHandsOnInstructions}\\ 
    Recognition variables & - &  See Section \ref{subsec:MethodsHandsOnInstructions}\\ 
    Wizard error  & Video representation, no variation between tests & Experiments were repeated when robot malfunctioned, occurrence: 1. \\ 
    \midrule
    Location   & Unknown, not asked due to privacy regulations & At the laboratory \\  
    Environmental constants   & Unknown & Setup as shown in Figure \ref{fig:figure2}  \\  
    Scenario   & As shown in video (see Supplementary Materials) & See additional information to scenario. \\  
    Iterative design process   & No & No \\ 
    Limitations   & \multicolumn{2}{c}{See Section \ref{sec:Limitations}}  \\  
  \bottomrule
\end{tabular}
\end{table}

\textbf{Reason for omitting the item "Experimental Component" from the Wizard presentation of Riek et al.:}
The hypothesis for the robot was to reproduce each participant's trajectory and create as little variation as possible. Thus, the robot’s software computed the trajectory planning. The points between the trajectories were programmed so that the overall task trajectory would be reproducible between participants. As Arboleda et al. indicated, robot behavior is perceived differently in virtual and real presentations \cite{Arboleda2021}. Therefore, a crucial aspect of the behavioral hypothesis was to compare the two presentations. Distinct reactions of the participants were not previously hypothesized, since the goal of this work was to gain insights into user expectations and behavior.

\textbf{Additional information on the item "Scenario":} The objects were placed in predefined positions within the robot’s workspace. All of the trajectories necessary for the automated task execution were hard-coded and could be selected by the experimenter by clicking on the program. The participant's safety was ensured by separating their workspace from the robot's.

\subsection{Online survey}
\label{subsec:MethodOnlineSurvey}
The online survey was chosen as the data collection tool because it provided a standardized way to present information and collect feedback. The goal was to obtain comprehensive feedback from PSMD, their family members, and medical professionals in order to evaluate the system's usability and accessibility. Providing an overview of the system to the stakeholder group via online survey and video minimized the logistical challenges described in Section \ref{subsec:BackgroundParticipation}. This allowed immobile individuals to participate using smart devices. Unlike a real-world demonstration, where the authors could only visit a few facilities, people living at home could participate. Additionally, the survey could be completed at the participant’s preferred time, which was convenient for those with complex schedules or who required assistance. Due to the anticipated limited availability of accessible computer controls, an extension was designed for family members and medical professionals to obtain more comprehensive feedback.
The purpose of the online survey was to answer the following questions: 
\begin{itemize}
    \item Are potential users able to use the system?
    \item Which tasks should be prioritized to maximize the robot's usefulness in everyday life?
    \item Were the participants satisfied with the presented system?
\end{itemize}

\subsubsection{Survey development and structure} 
The survey focused on investigating five topics, which are listed in Table \ref{tab:question_breakdown}. First, questions were asked about the participants' demographics to provide comparable and reusable baseline data. After the five topics were covered, participants were given the opportunity to provide feedback on the presented system and make general remarks. As stated in Section \ref{subsec:BackgroundParticipation}, perceptions of assistive devices can differ between end users and individuals in close contact with them, such as medical professionals and family members. Our goal was to receive a broad perspective on the prototype, which is why the survey varied between these three groups. Questions about the support situation were designed to evaluate the survey's accessibility. Poor accessibility can be indicated by how Q4(PSMD) is answered. A higher number of relatives and medical professionals who did not involve the person they care for can indicate time constraints. Furthermore, if everyone in Q4(relatives/friends) had stated that the person they care for had answered the survey, the answers would need to be weighted due to potential bias.

The topics of disabilities and assistive devices focus on whether the respondent is the intended audience. We implemented questions to reduce responses from people who misidentified their group affiliation. The topics also focus on whether the respondent is familiar with assistive devices, including robotic arms. The literature contains cases in which users reacted differently to devices when they had prior knowledge. This may affect their reaction to the presented system. The questions regarding the robot presentation were developed to identify signs of abandonment. Robotic arms are not widespread. Upon inquiry, two companies that sell assistive robotic arms approved as medical devices reported selling around 1,200 models in Germany, Switzerland, and Austria. Since this strongly affects the system’s perception, participants were asked if they had experience with robotic arms.

Lastly, task preferences were requested. As technology continues to evolve, the aim was to update the ongoing literature. As presented in Section \ref{subsec:BackgroundTaskPreference}, task preferences change quickly with the advent of new technology, such as smart home applications and hands-free smartphones. Some tasks, like using a fax machine, are no longer relevant. We asked participants to rank the importance of three tasks against each other.

Throughout the survey, we chose open-ended questions to receive rationale for the participants' answers to yes/no questions, such as reasoning for being unable to use the system. After initial development, the survey was discussed and refined.

\begin{table}[htbp] 
\caption{Breakdown of topics asked in the survey and related questions.}
\label{tab:question_breakdown}
\centering
\begin{tabular}{p{2.5cm}ccc}
\toprule
Question topic  & PSMD & Rel./Fr. & Med. prof.\\
\midrule
Support situation               & Q4+Q5   & Q4+Q5 & Q4\\
Disability \& assistive devices & Q6-Q8& Q6+Q7 &-\\
Robotic arm use                 & Q9-Q12&Q9+Q10&Q9+Q10\\
WoZ satisfaction                & Q13+Q14 &Q11+Q12+Q14+Q15&Q11+Q12+Q14+Q15\\
Task preferences                & Q15 & Q13 & Q5+Q13 \\
\bottomrule
\end{tabular}
\end{table}

\begin{figure}[ht]
  \centering
  \includegraphics[width=0.95\linewidth]{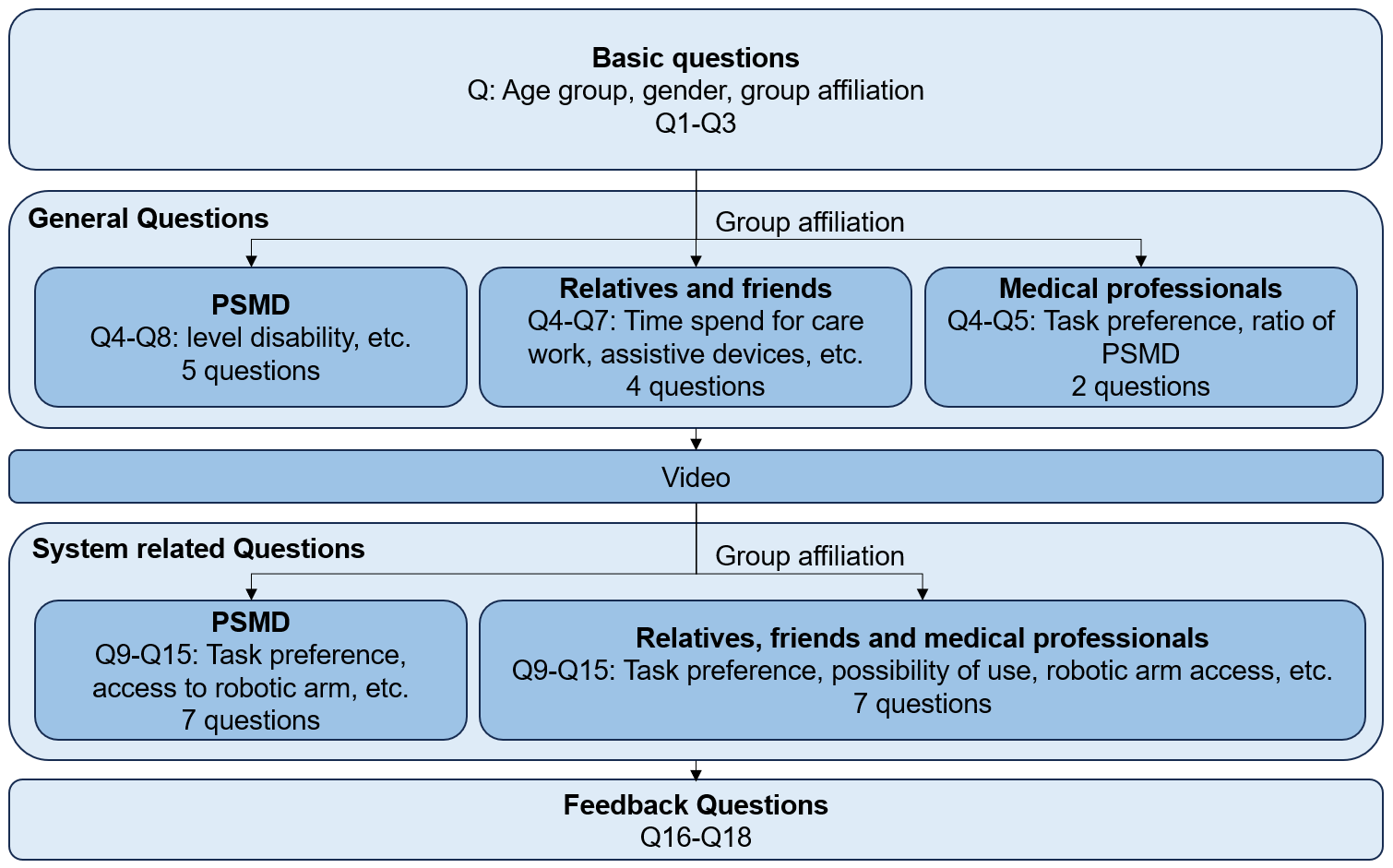}
  \caption{Distribution of questions asked for each group. Q outlines the context of the questions and the number of the questions for each category is given.}
  \Description{Distribution of questions asked for each group. Q outlines the context of questions asked and the number of the questions for each category is given.} 
  \label{fig:figure3}
\end{figure}

As a result, the online survey consisted of 31 questions, divided into three groups, as well as a video describing the eye-tracking control of the robotic arm, structured as shown in Figure \ref{fig:figure3}. In addition to the tasks, the video showed the calibration process. The exact questions asked are provided in Appendix \ref{app:appendix1}. All survey questions, along with the participants' responses, are available in the Supplementary Materials. The survey was designed using LimeSurvey.  

\subsubsection{Distribution method}
The survey was available online from September 18, 2023, to January 22, 2024. 32 organizations in Germany, Switzerland, and Austria were emailed to reach the stakeholder groups. Seven organizations, including FGQ (Paraplegic Support Association), LIS e.V. (Locked-In Syndrome Association), Diakonie (social services), Lebenshilfe (nonprofit association), and the Swiss Paraplegic Centre, shared the survey. The authors' private accounts and their institutions shared the survey a total of three times on social media. FGQ shared it in their private Facebook group, and the Swiss Paraplegic Centre posted about it in their private forum. 

\subsubsection{Participant instructions}
\label{subsubsec:MethodsOnlineSurveyInstructions}
First, participants were informed about the content and purpose of the study. Then, they were asked to give their consent to participate. Participants were then guided through the survey. Participants could skip questions at any time if they felt uncomfortable answering them. The video played when clicked on, and could be repeated if necessary. To offer different resolutions, participants could also open the video on YouTube.

\subsubsection{Evaluation method}
A frequency analysis was performed to evaluate users' perceptions of the system and their desired tasks. As introduced in Section \ref{subsec:BackgroundTaskPreference}, the tasks were categorized as follows: eating, drinking, picking up objects, environmental interactions, personal hygiene, and miscellaneous.

\subsection{Hands-on user study}
\label{subsec:MethodHandsOnStudy}
In addition to the stakeholder evaluation of the system, a Wizard of Oz design was employed to analyze real-world interactions. Evidence from the online survey suggests that participants can use the system based on their answers. However, challenges affecting the system's functionality may occur due to the eye tracker's insufficient accuracy or natural gaze behavior. Participants interacted with the robot to record gaze data, which was then used to evaluate hardware constraints. This study addresses the following questions:
\begin{itemize}
    \item Would the prototype eye-tracking controller work in the controlled environment of the lab?
    \item What technical difficulties arise from the design?
    \item Does perception vary between the hands-on study and the online survey?
\end{itemize}

\subsubsection{Procedure and participants}
\label{subsec:MethodsHandsOnInstructions}
To minimize the influence of learning effects or behavioral adaptation, 24 participants were involved, creating 24 emerging permutations in the task sequence. Each participant provided informed consent after the procedure was explained. They were asked to sit in a chair in front of the robot workspace. The eye-tracker was set up and calibrated as described in \cite{TobiiGlasses3}. 
Participants started the task by fixating on one of two objects in the drinking, eating, light-switching, and picking-and-placing tasks with gaze (see Figure \ref{fig:figure1}). Participants chose the objects according to their preference. The experimenter performed the role of the automation, as described in Section \ref{subsec:MethodsWOz}. 

Production and recognition variables were discussed prior to the test. Within the first second of selection, the wizard selected the corresponding task via the web interface. The robot’s reaction took between one and five seconds from the time the first reaction was recognized, including the wizard’s reaction time. As shown in Table 1, one wizard error occurred, causing the robot to stop shortly after the experiment started due to an incorrect mode selection. This experiment's results were discarded, and it was repeated with another participant to ensure naivety to the procedure and function of the Wizard of Oz setup.

Wizard training: The experimenter was informed of the user's task sequence and observed the participant's eye movements in real time on the monitor. As soon as the participant fixated on an object, the corresponding task started via the robot’s web interface.

\subsubsection{Data analysis}
Data were recorded using wearable eye-tracking glasses (Tobii, Tobii Pro Glasses 3, accuracy = 0.6$^{\circ}$). Gaze visualization, data analysis, and post-processing were performed using Tobii Pro Lab \cite{TobiiProLab}. In the hands-on study, the computer controlling the robot and visualizing the participant’s gaze was placed 2 m behind the participant so that the user was unaware of the experimenter’s role. Evaluating the areas of interest (AoI) and dwell time is necessary to estimate the need for error minimization methods and determine constraints for object recognition algorithms. Along with this technical evaluation of the system, we analyzed the perception of the non-disabled participants. 
In the Tobii Pro Lab events, Times of Interest (ToI), gaze movements during object selection, and Areas of Interest (AoI) were defined and labeled on the data. Tobii Pro Lab calculated metrics such as dwell times, which describe the total fixation time of an AoI relative to the ToI and the number of missed hits, and exported them for further analysis. The technical evaluation was performed based on these parameters. 

Interpreting gaze data provides insights into the factors mentioned in Section \ref{subsec:BackgroundSharedControl}. ToI, AoI, and dwell time can be used to describe the scene based on visual attention. In order to facilitate the understanding of such terms, definitions according to Holmqvist and Nyström \cite{Holmqvist2017} and Tobii \cite{TobiiProLab} are presented below. 
\begin{itemize}
    \item Event: In Tobii Pro Lab, describes the timestamp when an event occurs. In this case, it marks the start and end times of a task.
    \item Time of Interest (ToI): The interval between two related events.
    \item Area of Interest (AoI): A polygon that corresponds to the outline of an object. The resulting area describes a visual stimulus.
    \item AoI hit: The entry of the tracked gaze into an AoI.  
    \item Dwell Time: The time that the gaze remains within an AoI after the initial entry. In this study, the gaze was allowed to exit and re-enter the AoI. Dwell time is the sum of time while the gaze stays within an AoI. 
    \item Type of eye movement: In this study, fixations and saccades were of primary interest. Fixations refer to the time that the eyes remain stable and focused on a specific point or area of interest. They were not distinguished from smooth pursuit, which occurs when the head moves while the gaze remains fixed on a point in the scene. Saccades are rapid changes in gaze direction. They connect fixations. Other eye movements, such as vergence movements and microsaccades, were not investigated. Further information on gaze movement can be found in Holmqvist and Nyström \cite{Holmqvist2017}.
\end{itemize}

\section{Results}
\subsection{Online survey}
\label{subsec:ResultsOnlineSurvey}

Of the 71 participants, 36 surveys with no responses (N/A) had to be excluded. 19 participants completed the survey, and 15 partially completed surveys were included when they answered at least Q1 to Q12 (referring to the question indicators stated in Appendix \ref{app:appendix1}), due to their given insights into daily life. Therefore, the analysis used 34 complete and partially completed surveys. We assume that the large number of empty surveys resulted from people skipping through the survey out of curiosity. Additionally, some institutions reported that they conducted an internal ethical evaluation before distributing the survey to patients and staff. These surveys were automatically saved, even if they were not finished.  

The group distribution of the completed surveys was 10 PSMD, 3 family members or friends (fam./fr.), and 6 medical professionals (med. prof.). In comparison, the in the incomplete surveys 4 PSMD, 6 fam./fr., and 5 medical professionals participated. The demographics can be found in Table \ref{tab:tab2}, which summarizes the results of Q1 to Q3. The general and system-related questions varied by group to identify different viewpoints on the system as shown in Figure \ref{fig:figure3}. Table \ref{tab:question_breakdown} shows the distribution of questions related to the corresponding topic. The results of Q4 to Q15 will be presented in the following topics.

\subsubsection{Support situation}
For this topic, questions were asked to evaluate the participants' personal situations and whether there were any obstacles to filling out the survey. Only two of the PSMD group received assistance filling out the survey (Q4(PSMD)). The reason was not stated. Due to limitations in hand and upper body movement, and a lack of access to suitable assistive devices, it can be assumed that the assistance was necessary (Q8, Q9). Relatives and friends only stated once that the person they cared for filled out the survey. Therefore, the risk of bias due to conferring between participants is low.

\subsubsection{Level of disability and assistive devices} 
Q6 revealed that two participants experienced variability in their disability-related limitations from day to day, underscoring the importance of considering individuals with episodic disabilities. These individuals may particularly benefit from multimodal control approaches that can adapt to varying abilities, as discussed by Bhattacharjee et al. and Nanavati et al. \cite{Bhattacharjee2020, Nanavati2024b}. Regarding Q8, participants differed greatly in their motor abilities and, in most cases, their level of disability remained unchanged (Q7: 12 out of 15). In the other three cases, the level of disability improved, which could indicate stroke rehabilitation. This emphasizes that, for this group, the presented control method may only be applicable for a short time before switching to other control inputs, such as a joystick. This emphasizes the importance of systems that are easy to learn and apply, assisting in short recovery times. Relatives and friends stated that three participants had access to assistive devices, in the form of a patient lift and a repositioning aid.

\subsubsection{Robotic arm use}
Of all the participants asked, only one stated that she had already used a robotic arm controlled by a chin switch. In the context of Q7(PSMD) and Q8(PSMD), it may be that the level of disability of the other participants does not qualify them to receive a robotic arm from healthcare providers, or they may receive assistance from caregivers or family, which is more common in Germany, Switzerland, and Austria. Approval for a medical device by healthcare providers in Germany can take several months or, in rare cases, years. This indicates that the use of robots is uncommon and can greatly bias the participants' perception of the system due to its novelty.

\subsubsection{WoZ satisfaction}
After viewing the video, the participants were asked if they or the person they cared for would be able to use the system. More than half of the participants confirmed this (17 out of 34; Q13 - PSMD, Q14 - Rel./Fr. and Med. Prof.). Only three participants stated the opposite. No reasons were given. Medical professionals and family members were also asked if they were satisfied with the setup process (Q11 - Rel./Fr. and Med. Prof.). Five stated they were satisfied, while three were not. One reason given was the amount of time it took to set up. It might also be influenced by the desire for more information about the system. As Figure 4 shows, the high number of N/A answers in the partially completed surveys occurred after the video. This may be because participants were unable to return to the website after watching the video. Due to missing feedback, this is only a presumption.

\begin{figure}[ht] 
  \centering
  \includegraphics[width=0.95\linewidth]{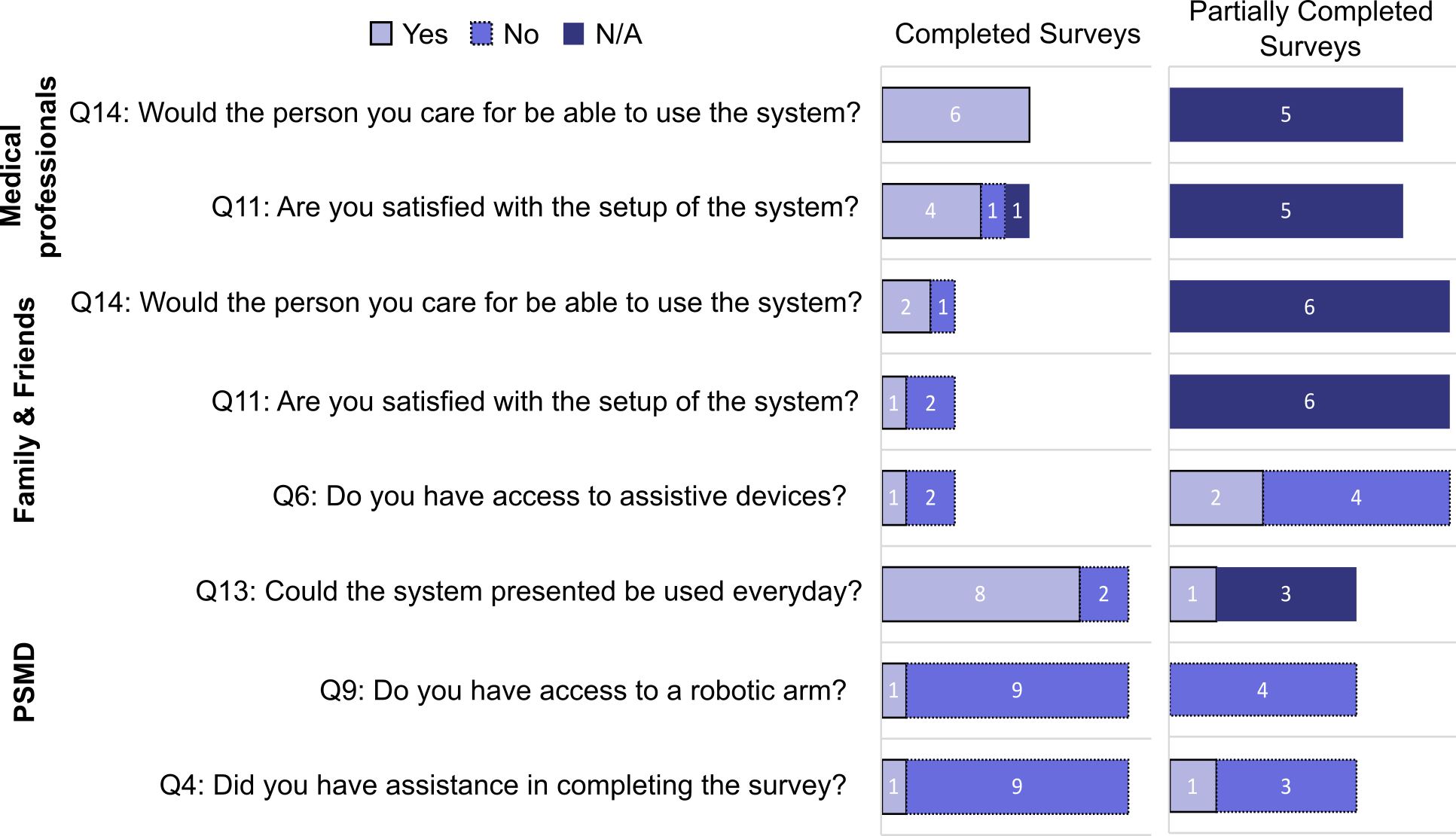}
  \caption{Distribution of responses from online survey participants. Q4, Q9 and Q13 were answered by PSMD, Q6, Q11 and Q14 by family and friends and medical professionals.} 
  \Description{This bar chart shows the distribution of responses of online survey respondents. Eight questions are presented that highlight the most interesting findings. PSMD were asked the following three questions: First question: Did you have assistance in completing the survey? Responses: fully completed surveys 2 yes, 12 no, 0 not answered. Second question: Do you have access to a robotic arm? Responses: fully completed surveys 1 yes, 13 no, 0 not answered. Third question: Could the system presented be used everyday? Responses: fully completed surveys 9 yes, 2 no, 3 not answered. Family members were asked the following three questions: First question: Do you have access to assistive devices? Responses: fully completed surveys 3 yes, 6 no, 0 not answered. Second question: Are you satisfied with the setup of the system? Responses: fully completed surveys 1 yes, 2 no, 6 not answered. Third question: Would the people you care for be able to use the system? Responses: fully completed surveys 2 yes, 1 no, 6 not answered. Medical professionals were asked the following two questions: First question: Are you satisfied with the setup of the system? Responses: fully completed surveys 4 yes, 1 no, 6 not answered. Second question: Would the person you care for be able to use the system? Responses: fully completed surveys 6 yes, 0 no, 5 not answered.}
  \label{fig:figure4}
\end{figure}

\subsubsection{Task preference}
\begin{figure}[ht]
  \centering
  \includegraphics[width=0.6\linewidth]{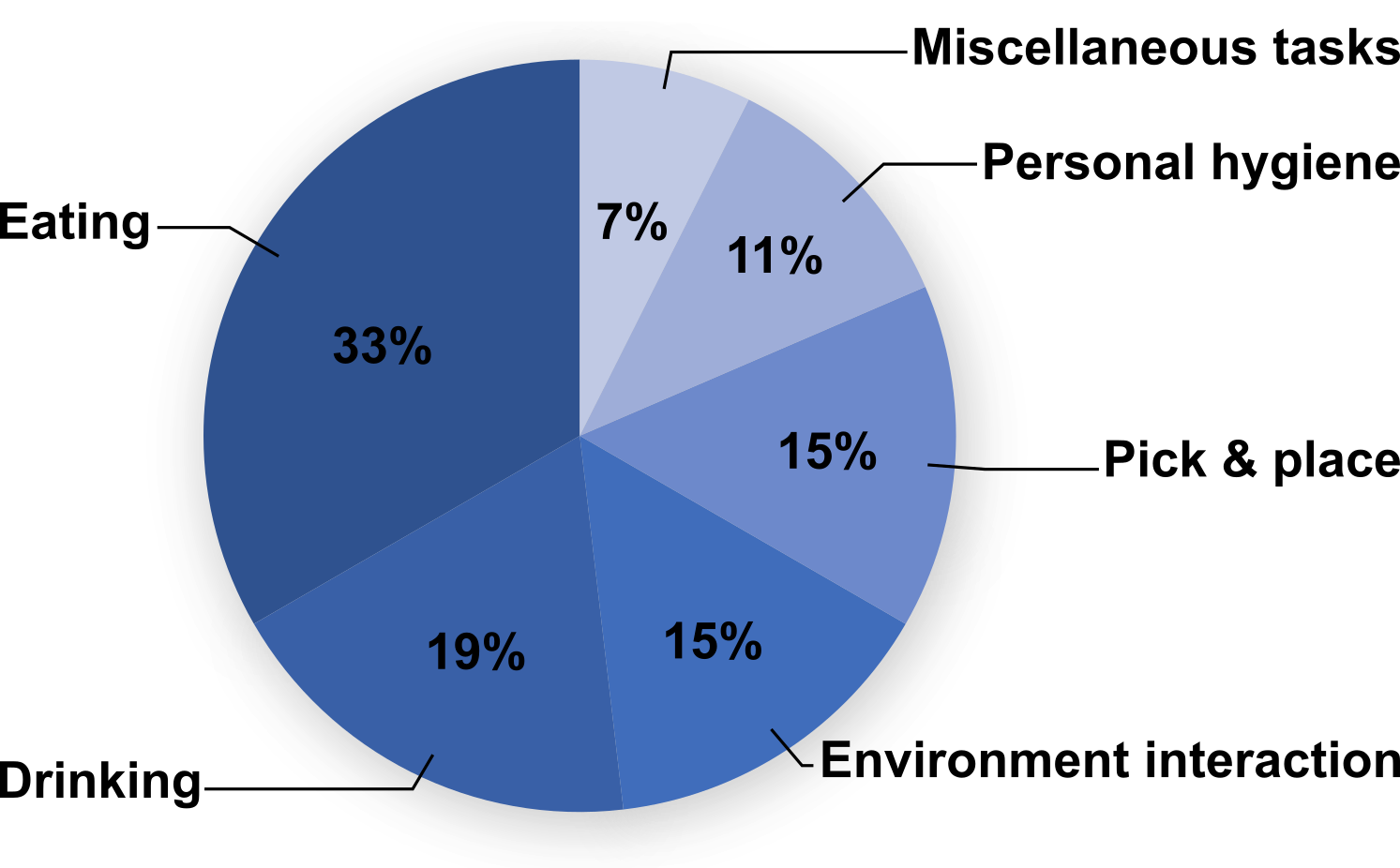} 
  \caption{Classification of all mentioned tasks in relation to the presented system. 28 tasks were classified into six categories containing the tasks mentioned by the participants.}
  \Description{This pie chart shows the classification of all mentioned tasks in relation to the presented system. 28 tasks were grouped into six categories containing the tasks mentioned by the participants and are shown in percentage. The largest category is nutrition with 52\%, the second largest category is daily tasks with 19\%, the third is mobility with 11\%, the fourth is two categories with 7\% each described by the terms personal hygiene and well-being. Finally, tasks classified as communication are mentioned with 4\%.}
  \label{fig:figure5}
\end{figure}

Participants were asked what tasks they would prioritize to accomplish with the system. The group of PSMD gave a total of 14 responses: Eating (3 mentions), drinking (2 mentions), reaching for cutlery, pushing buttons (elevator or electric door), applying makeup or brushing their hair, cleaning, scratching, opening a bottle, and lifting objects of various weights from the floor (2 mentions). Family members and medical professionals were asked this question twice. First, they were asked to identify situations in which a robotic system could assist with daily tasks, while they were unaware of the system demonstration. Second, they were asked about the presented system. Thirteen people responded to the first question. They rated the following tasks as useful: Making the bed, brushing teeth, assisting with eating, thrombosis prevention, putting on socks, documentation, going to the bathroom, doing laundry, cleaning (2 mentions), assisting with walking, assisting with doctor's appointments and correspondence, silent alarm for service calls, and making phone calls. Fourteen responses were given to the second question. Eating (5 mentions), drinking, turning on lights, operating the aforementioned lift system, brushing teeth, preparing a drink, opening doors, lifting objects (2 mentions), making phone calls, and moving objects. Both questions demonstrate a high potential for robotic systems and digitization. The first focuses on assistance with household tasks and communication, while the second focuses on user-centered tasks. Figure \ref{fig:figure5} shows the distribution of task preferences mentioned by the participants. 

\subsection{Hands-on study}
\label{subsec:ResultsRealWorldStudy}
\subsubsection{Gaze behavior analysis} 
Analyses of dwell times in relation to object selection events were used to determine the technical requirements for the eye-tracking controller. Assumptions were investigated, such as the correlation between object size and dwell time duration.

\begin{figure}[ht]
  \centering
  \includegraphics[width=\linewidth]{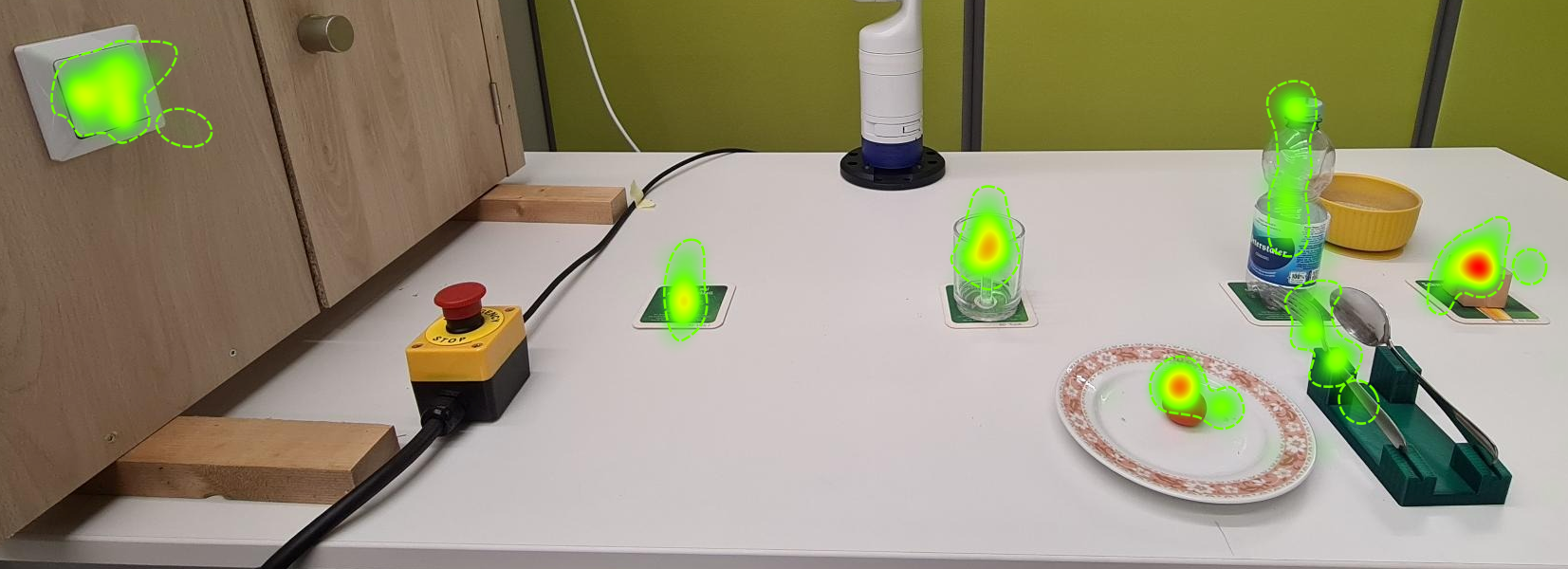}
  \caption{Heat map visualizing the attention of participants during task selection (absolute duration), mapped to the snapshot of the experimental setup. Red areas indicate frequent fixations, green less fixations. Areas without coloring indicate no fixations within the observed intervals. Dashed outlines were added manually to better distinguish the edges of the hotspots.} 
  \Description{The figure shows the experimental setup shown in Figure \ref{fig:figure1}. It is overlaid with a heat map visualizing the attention of all participants during task selection (absolute duration). The heat map is visible on the objects and, in some cases, the area around the objects. Task 1: Three hotspots of low and medium intensity are shown on the light switch. There are few fixations outside the light switch towards the center of the experimental setup. Task 2 - Pick and place: The block shows a high intensity hotspot at the top of the block in the left corner toward the center of the experimental setup. The hotspot extends beyond the boundary of the object. A few fixations to the right of the block are visible in the vicinity of the block. The platform shows an elliptical hotspot perpendicular to the height (y-axis) of the figure with low, medium, and high intensity. Few fixations were found outside the platform above the object. Task 3 - Filling a cup: The cup's high-intensity hotspot is centered on the top part of the handle and extends slightly beyond the object's outline. The bottle has a vertically elongated hotspot that starts at a third of the height of the bottle from the bottom and continues to the bottle cap, slightly exceeding the objects outlines. Although only low intensities are shown, the highest intensity is at the bottle cap. Task 4 - Picking a food item: The hotspot on the food item is off-center from the food item itself and above the food item. The hotspot is about the same size as the food item and covers approximately half of the food item itself. A second low-intensity hotspot is shown to the right of the food item, between the food item and the fork. The fork has two hotspots at the base of the fork tines and slightly below in 1/6 of the fork handle. Both hotspots extend across the width of the fork.}
  \label{fig:figure6} 
\end{figure}

Table \ref{tab:tab1} shows the distribution of participants' object selection choices and dwell time information. A visual attention map is shown in Figure \ref{fig:figure6}. The chosen interval was 1,043\,ms before the robot started moving, based on the time steps made by the evaluation software. Both data in Table \ref{tab:tab1} and in Figure \ref{fig:figure6} represent the intervals. During this time period, events such as searching for the object or deciding which object to select were excluded. Participants were instructed to focus solely on the selected object, to investigate whether natural gaze behavior exhibits gaze cues that hinder implementation or result in insufficient accuracy. The average duration of the test was 2\,min, 41\,sec (SD = 9.8 sec).  

The average dwell time was calculated to determine the maximum limit for the controller's dwell time, including the time from the start of fixation until the robot moved. In this setting, the longest possible dwell time for the participant’s gaze was found to be 5.612\,seconds. During this time, the gaze did not leave the area of interest (AoI). This indicates that dwell time triggers can be set to long durations. However, the following limitations show that such long dwell times are rare.

\begin{figure}[ht]
  \centering
  \includegraphics[width=0.9\linewidth]{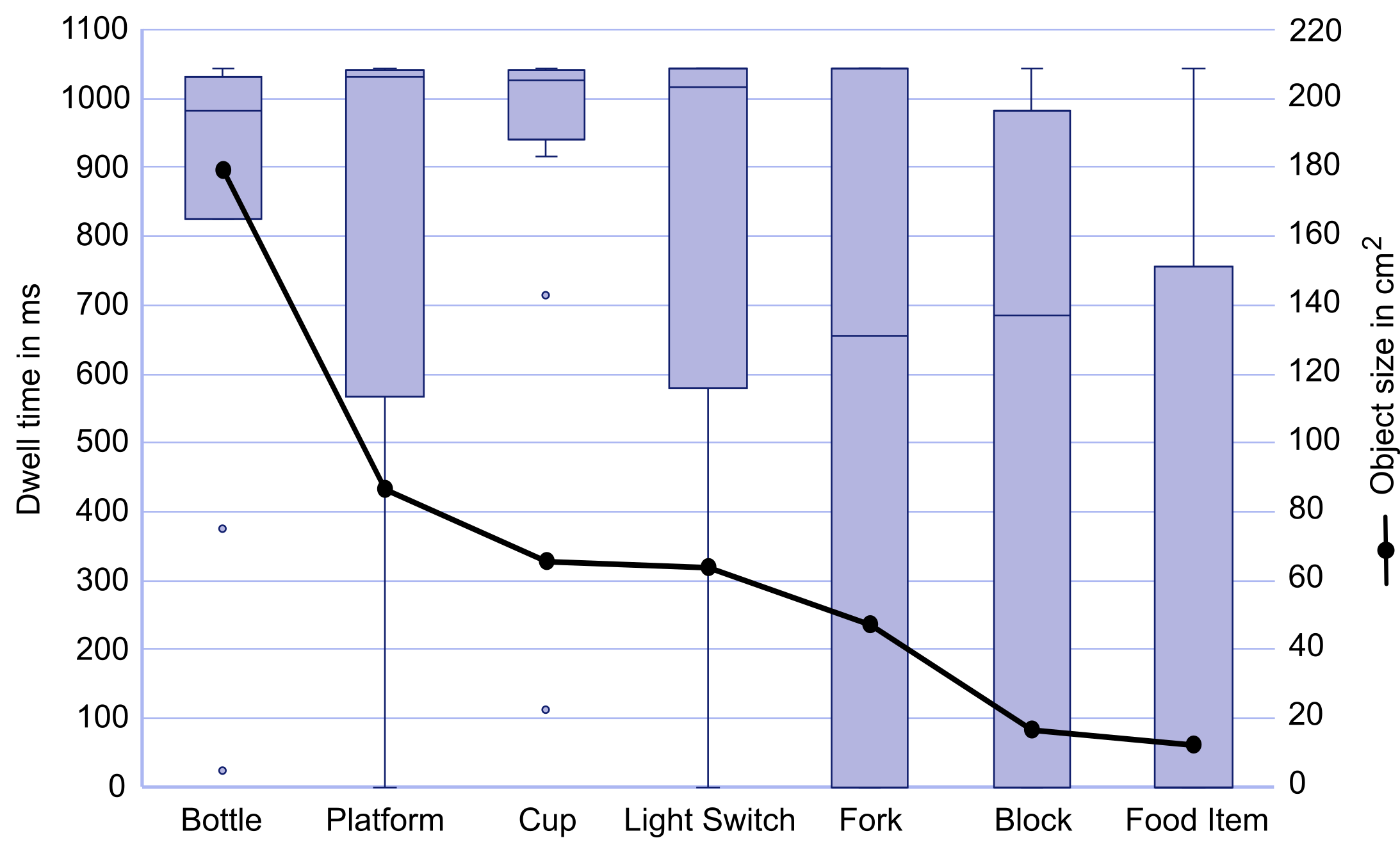}  
  \caption{Correlation between object size, missed hits and dwell time for the selectable objects. Dwell time is given as the total duration of gaze within the objects AoI 1043\,ms before the robot started moving. The line in the box plots marks the median, confidence intervals are given at 95\%.}
  \Description{This graph shows the correlation between object size, missed hits and dwell time for the selectable objects. Dwell time is given as the total duration of gaze within the objects AoI 1043\,ms before the robot started moving and visualized as a box plot. In the following the average, median whiskers and outliers are described for each object. 
  Bottle: Median = 983\,ms, SD = 826\,ms to 1043\,ms, 2 outliers at 375\,ms and 24\,ms. 
  Cup: Median = 1027\,ms, SD = 917\,ms to 1040\,ms, 2 outliers at 715\,ms and 113\,ms.
  Fork: Median = 654\,ms, SD = 0\,ms to 1043\,ms, no outliers.
  Food Item: Median = 0\,ms, SD = 0\,ms to 1040\,ms, no outliers.
  Platform: Median = 1031\,ms, SD = 0\,ms to 1043\,ms, no outliers.
  Block: Median = 684\,ms, SD = 0\,ms to 1043\,ms, no outliers.
  Light switch: Median = 1016\,ms, SD = 0\,ms to 1043\,ms, no outliers.
  Two line plots representing object size and missed hits were superimposed on the box plot. Line plot results for the object size in square meter: Bottle: 0.018, cup: 0.00657, fork: 0.00475, food item: 0.001225, platform: 0.008649, block: 0.0016, light switch: 0.0064. Line plot results for missed hits of the AoI of each object: Bottle: 0, cup: 0, fork: 5, food item: 7, platform: 1, block: 5, light switch: 4. }
  \label{fig:figure7}
\end{figure}

In this study, we assumed that dwell time would be longer for larger objects due to larger AoIs. Accuracy errors would have less impact on AoI interactions than in the small AoIs. The correlation between object sizes and corresponding dwell times was measured and is given in Figure \ref{fig:figure7}. The dwell time of the bottle has a higher standard deviation than the cup, implying more occurrences of longer dwell times. As seen in Table \ref{tab:tab1}, this is also visible in the dwell times above 500\,ms. Reasons why the correlation may not apply are the fixation of the bottle cap, which is closer to the edge of the AoI than the cup handle. The gaze is more likely to leave the AoI, resulting in lower dwell times, which indicates the influence of \textit{user intention}. In comparison, the fork and the block have similar average dwell times and standard deviations, despite their different sizes. We extended our initial hypothesis to include the influence of \textit{geometry of the object} on the dwell time. We tested this by evaluating missed hits and gaze behavior with a fork handle that was less than 1\,cm wide. When accuracy errors occur, it is more likely that gaze is tracked outside the fork's AoI.  

\begin{table} [ht]
  \caption{Information about distribution of selection by participants, dwell times and missed AoI hits.}
  \label{tab:tab1}
  \begin{tabular}{cccc}
    \toprule
    Object & Number of selections & Occurrences of dwell time over 500\,ms & Missed AoI hits\\
    \midrule
    Bottle      & 11 & 82\% & 0 \\
    Platform    & 7  & 87\% & 1 \\
    Cup         & 13 & 92\% & 0 \\
    Light switch& 24 & 75\% & 4 \\
    Fork        & 9  & 58\% & 5 \\
    Block       & 14 & 56\% & 5 \\
    Food item   & 11 & 38\% & 7 \\   
  \bottomrule
\end{tabular}
\end{table}

\begin{figure}[ht]
  \centering
  \includegraphics[width=1.0\linewidth]{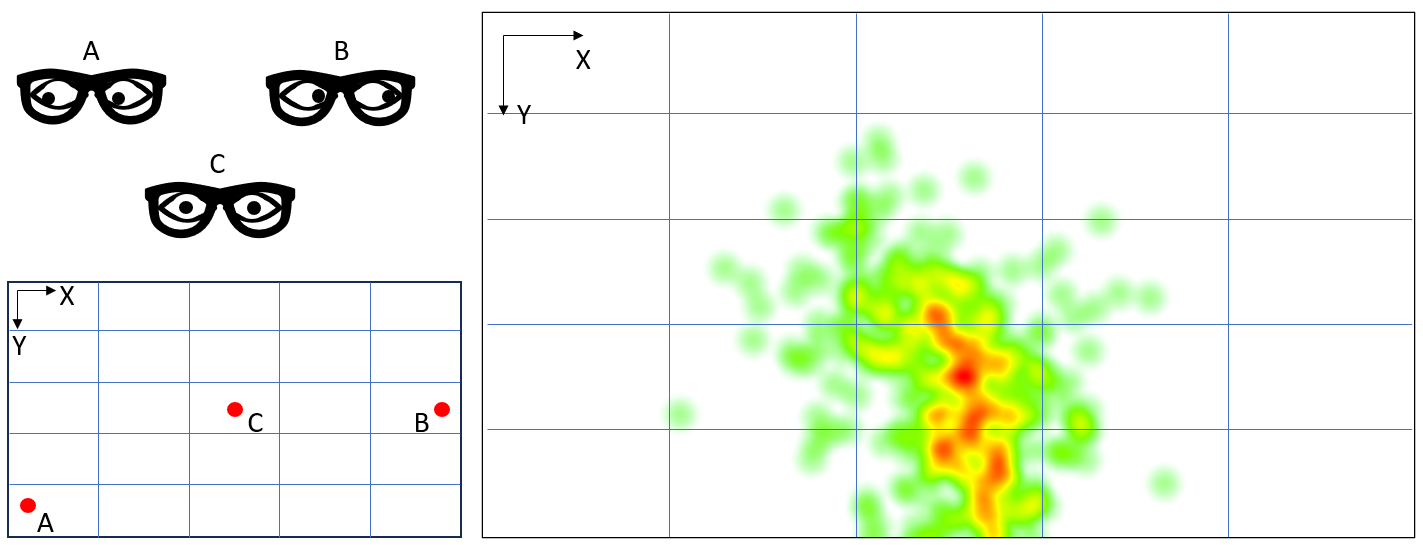}
  \caption{Left: Exemplary representation of the viewing directions to explain the resulting heat map. Visualization of the viewing point in relation to the visual field of the scene camera. Right: Heat map of gaze position during task selection. The field of view is divided into 5 by 5 fields indicating a 5-section coordinate map used to determine the correlation between dwell time, gaze location, and missed AoI hits. Gaze direction was mostly centered and downwards while task selection.}
  \Description{This figure shows the heat map visualization of  gaze position in the field of view of the eye-tracking camera during task selection. The field of view is divided into 5 by 5 fields indicating a 5-section coordinate map used to determine the correlation between dwell time, gaze location, and missed AoI hits. Starting from the upper left corner, the fields are labeled as follows: A1 is the upper left corner, A5 is the upper right corner, E1 is the lower left corner, and E5 is the lower right corner. The highest intensity of the participants' attention is found in C3, D3, and E3. Medium intensity is also found in C2, D2, E2, D4, and D5. Low intensity is found in all fields except row A and columns 1 and 5. Therefore, the participants' gaze is centered in the middle of the field of view in the x-direction and mostly in the middle and lower part in the y-direction.}
  \label{fig:figure8}
\end{figure}

Examining the low dwell times of the food item placed closest to the user revealed the influence of \textit{gaze direction} on the dwell time. Looking at the gaze distribution in relation to the scene camera's coordinate system reveals that participants tended to move their heads to center their gaze when looking at objects to the right or left of them, as shown in Figure \ref{fig:figure8}. In contrast, participants tended not to move their heads when looking down. The detected gaze location was less accurate due to the greater scatter closer to the eye-tracker's periphery. Physiological factors, such as shadowed pupils by the eyelashes, cannot be excluded to impact dwell time. This resulted in seven missed AoI hits for the food item. 

Significant correlations were investigated between dwell time, object size, and gaze direction towards the object. The Shapiro-Wilk test (H\,=\,0.771, df\,=\,97, p\,=\,<0.001) and the Levene test (H\,=\,2.658, df1\,=\,6, df2\,=\,90, p\,=\,0.02) showed no normal distribution or homogeneity. A Kruskal-Wallis test was applied and showed significant differences ($H(6)=17.442$, $p<0.008$). Post-hoc analysis using Dunn's test with Bonferroni correction revealed significant differences ($p<0.05$) between the following pairs: food item and bottle, food item and platform, food item and light switch, food item and cup, and block and light switch. These results indicate that the measured data for the food item are significantly different from the other objects. 
Spearman's $\rho$ was used to analyze the correlation between object size, object position, and dwell time. In Figure \ref{fig:figure8}, gaze location was evaluated by applying a coordinate map visualized by the equally distributed sectors. The analysis showed that dwell time and object size ($\rho=0.373$, $p<0.001$) were positively correlated, as well as dwell time and offset in y-direction ($\rho=0.263$, $p=0.009$). The correlation between the offset values in y-direction and x-direction is weaker and does not correlate with dwell time ($\rho=0.160$, $p=0.116$). This indicates that the assumption that the dwell time is influenced by \textit{gaze location} towards the object and the \textit{object size} is correct.

\subsubsection{Perception of hands-on application}
In general, the hands-on study revealed satisfaction with the system. In the unstructured interview, the participants' comments were evaluated based on whether they were positive or negative regarding the robot's performance. The results were as follows: nine positive comments, three negative comments, nine comments that were both positive and negative, and three comments that were neither positive nor negative. All statements can be found in Appendix \ref{app:appendix2}.
Differences were found between the responses of people who regularly work with robotic arms (the prior-experience group) and those who have no experience with robotic arms and robots in general (the non-prior-experience group). The non-prior-experience group gave mostly positive comments and focused on describing the tasks, while the prior-experience group focused on robot kinematics, robot behavior, and user safety.

The non-expert group reported feeling interested, entertained, satisfied, and impressed. At the same time, they reported feeling confused and uncertain. The task design itself caused confusion because participants were unsure how to react to the glass presented by the robot. Considering a real-life application of pouring a drink, one participant expressed fear of overfilling the glass. 
Regarding the robotic system, participants were satisfied with good reaction time and intuitive control. Negative comments concerned the velocity, which created ambivalent reactions. Some participants stated the velocity was just right, and others saying it was too slow. Importantly, users reported feeling safer because the robot's movements were limited to the edge of the table. The fork movement toward the participant had an unsettling effect, as they reported that the fork was not close enough to eat, or at the right height for their mouth. 

Comments from the prior-experience group were mostly negative about the robot's trajectory and the measurement setup. For instance, the gripper could potentially crush someone's hand if the user reached into the setup due to the lack of pressure sensors. The placement of the emergency stop was a concern for the well-being of future users, due to its distance from the participant. They also mentioned the robot's shaky movements, non-optimized trajectories, and one instance where the robot accidentally touched the bottle while moving to the next object. The perceived motor vibration was uncomfortable and the sound of the dropped fork was too loud. Reaction time and object recognition were mentioned as positive, implying that the Wizard of Oz design was convincing. 

Additionally, two participants from both groups stated that they were uncertain whether the robot would stop the task if they looked away or at other objects while the robot was performing it, a concern that must be addressed in the control design.

\section{Discussion}

In Section \ref{subsec:DiscussionUsability}, two of the three usability questions and in Section \ref{subsec:DiscussionFunctionality}, two of the three functionality questions are answered. In Section \ref{subsec:DiscussionPerception}, the last question of both studies will be discussed, which combines the perception of the online study and hands-on study. Feature improvements to the initial design are presented.

\subsection{Usability}
\label{subsec:DiscussionUsability}
\subsubsection{Are potential users able to use the system?}
Yes, the survey showed in Q13(PSMD) and Q14(Fam./Fr. and Med. prof.) (Figure \ref{fig:figure4}) that most of the participants are able to use the system based on the video presentation (Answers: 17 yes, 3 no, 14 N/A). The high amount of N/A results were the result of including partially filled out surveys.  

The limiting factors that prevent the system from being used every day were not stated. In the open-ended questions, participants stated that they would like more information. They also said that they would like more information about the system to better evaluate its maintenance and function. These requests did not imply that the participants would change their minds after receiving the additional information. One participant stated, “I wonder if the system can be used by people with intellectual disabilities due to their reduced cognition”. This highlights the importance of considering various impacts of disabilities during the design process to make such systems accessible to a wide range of users. It again highlights the necessity of involving the community in the design process.  

Deeper issues, such as how the control works and whether wrong selections can occur due to natural gaze behavior, were not questioned by the participants, contrasting the results from the hands-on study. Section \ref{subsec:DiscussionPerception} shows participants' concerns about interacting directly with the system, highlighting the benefit of dual investigations in robot perception.

\subsubsection{Which tasks should be prioritized to maximize the robot's usefulness in everyday life?} 
The results, shown in Figure \ref{fig:figure5}, the most requested tasks are assisting while eating or drinking, followed by environmental interaction, which includes turning lights on and off, pressing buttons, opening doors, and picking and placing objects. Personal hygiene consisted of a variety of different tasks, including doing makeup and hair, brushing teeth, and cleaning. Miscellaneous tasks included making phone calls and scratching themselves.

Brushing one's hair, applying makeup, or scratching are daily tasks that require close human-robot interaction while ensuring the safety of the user. The shared control approach presented would require face and body recognition if the user is unable to move their head to a certain position. Ethical questions arise when personal data is recorded, including physical and physiological appearance. Current discussions in this area suggest that the user must be informed about the system functions and the personal data collected. This would allow the user to decide whether the privacy limitations or the benefits of the system outweigh \cite{FoschVillaronga2020}. The aim must be to present this information in simple language so that people with physical and cognitive disabilities can understand it and make decisions.

The prioritized tasks and the tasks realized from the literature included in Section \ref{subsec:BackgroundTaskPreference} were summarized. A comparison of the results with the findings in the literature shows partial agreement. Table \ref{tab:overviewtasks} compares findings from the literature with our results. In the "Prioritization" column, we summarized the works of Beer et al., Nanavati et al., Petrich et al., Smarr et al., Stanger et al., Chang et al., and Jardón et al. \cite{Beer2011, Petrich2022, Nanavati2024b, Smarr2012, Stanger1994, Chang2003, Jardon2012}. These works present or review task prioritization in activities of daily living (ADLs), involving either elderly people or people with disabilities. It is important to acknowledge that these works include varying participant groups, as well as varying robot types. To the best of the authors' knowledge, no review or work solely focuses on PSMD and robotic arms. The "Integration" column summarizes reviews and individual studies concerning the execution of tasks with assistive robots. It summarizes the outcomes of the following works, which were also presented in Section \ref{subsec:BackgroundTaskPreference} \cite{Nanavati2024b, Hagengruber2025, Dragan2013, Huang2016, Zhang2025, Bien2003, Canal2021, Bhattacharjee2020}. Some tasks were excluded from the literature when they could not be performed with a robotic arm (e.g., getting around or into bed, going out, toileting, managing money, leisure and recreation, and shopping), as well as outdated tasks (e.g., removing paper from printers and fax machines).

\begin{table}[ht]
    \centering
    \caption{Overview of tasks mentioned by participants and representation within literature. Tasks mentioned by users are presented to their according categories in Figure \ref{fig:figure5}. Fit describes if the task can be accomplished without additional assistive devices to a robotic arm (triangle shows partial accomplishment). The comparison is done between our results, task prioritization found in other works and integrated tasks in existing works. }
    \Description{The table presents an overview of daily-life task categories, specific tasks within each category, and how these tasks are represented in the collected user responses and in prior literature on assistive robotic arms. Categories include Eating, Drinking, Environmental Interaction, Personal Hygiene, and Miscellaneous Tasks. For each task, the table indicates whether it can be performed by a robotic arm without additional assistive devices, and compares three aspects: how frequently the task was mentioned by participants in the hands-on study, how strongly participants in related literature studies preferred the task, and how strong the research interest in realizing the task in existing assistive robotic systems is. 
    Eating and drinking are both considered feasible with a robotic arm alone and are frequently mentioned by users, commonly prioritized in literature, and often integrated in existing systems. Environmental interaction tasks include preparing meals, using switches and buttons, opening doors and drawers, pick-and-place actions, and carrying objects. All of these tasks can be accomplished with a robotic arm alone. All tasks except preparing meals are mentioned by participants in this study. They are highly prioritized by participants in other studies, and commonly integrated in existing systems, with Switches, Buttons and carrying objects showing slightly less frequent integration.
    Personal hygiene tasks include makeup or shaving, combing hair, brushing teeth, dressing, and bathing or grooming. All are feasible with a robotic arm alone, except bathing and grooming. Bathing and grooming are only partially feasible without additional assistive devices. They mentioned by participants in this study except Dressing and Bathing which also represents the results from other literature. They are less commonly integrated into existing systems. 
    Miscellaneous tasks include using phones or electronics, taking medicine, doing household chores, and scratching. Household chores are feasible, mentioned in this study and literature and frequently integrated in existing systems. Using electronics and scratching are only partially feasible without additional devices. Taking medicine is feasible but less frequently mentioned and less often integrated in existing systems.}
    \label{tab:overviewtasks}
    \begin{tabular}{p{2.2cm}p{3.5cm}cccc}
    \toprule
    Category        &	Tasks	         & Fit	  & Our results       &	Prioritization      &	Integration \\
    \midrule
    Eating	        &Eating              &	$\checkmark$  &	$\nearrow$	      & $\rightarrow$          &	$\uparrow$ \\
    \midrule
    Drinking	    &Drinking            &	$\checkmark$	&$\nearrow$	         &$\nearrow$	          &$\uparrow$ \\
    \midrule
    Environmental   &Preparing meals	&$\checkmark$&		          &$\uparrow$	             &$\uparrow$ \\
    Interaction                &	Switches, Buttons &	$\checkmark$   &	$\nearrow$	     &$\uparrow$	       &$\rightarrow$ \\
	                  &Doors, drawers   &	$\checkmark$	&$\nearrow$	   &$\uparrow$	           &$\uparrow$ \\
	                &Pick and place	     &$\checkmark$     &	$\nearrow$	     &$\uparrow$	       &$\uparrow$ \\
	                &Carrying objects     &	$\checkmark$    &	$\nearrow$	         &$\uparrow$            &	$\nearrow$ \\
                    \midrule
    Personal  & Makeup, shaving&	$\checkmark$&	$\nearrow$	&$\uparrow$	&$\rightarrow$ \\
	hygiene                &Combing hair        &	$\checkmark$&	$\nearrow$	&$\nearrow$	&$\searrow$ \\
	                &Brushing teeth      &	$\checkmark$&	$\nearrow$	&$\nearrow$	&$\searrow$ \\
	                &Dressing            &	$\checkmark$&		             &$\rightarrow$	       &  $\rightarrow$ \\
	                &Bathing, Grooming    &	$\triangle$	&	&$\rightarrow$	&$\nearrow$ \\
                    \midrule
    Miscellaneous            &Using phone, electronics&	$\triangle$&	$\nearrow$	&$\rightarrow$ &	$\rightarrow$ \\
	tasks                &Taking medicine     &	$\checkmark$&		&$\rightarrow$&	$\searrow$ \\
    	                &House choirs &	$\checkmark$&	$\nearrow$&	$\uparrow$	&$\uparrow$ \\

	                &Scratching	         &$\triangle$&	$\nearrow$	&$\nearrow$&	$\searrow$ \\
                    \bottomrule
    \end{tabular}
\end{table}

Tasks such as preparing meals, dressing, bathing, grooming, and taking medicine were not mentioned by the participants. This may be because these tasks are comparatively less important to end users, as found in the literature. In contrast, preparing meals is highly prioritized and researched in other studies. Depending on their hobbies or the type of personal assistance they receive, this task may not have been favored by the small group of participants. The insight into the eating task is interesting. While most participants stated that they strongly preferred to accomplish this task independently, Nanavati et al. found that, compared to other ADLs and IADLs, this task was the least prioritized \cite{Nanavati2024b}. However, other sources, such as Petrich et al. and Stanger et al., found it to be highly prioritized \cite{Stanger1994, Petrich2022}. This discrepancy may be due to the fact that Nanavati et al.'s work reviewed different populations, including people with visual impairments and the elderly, who may be able to eat without assistance. 

Comparing works that address these tasks reveals gaps in interacting with switches and buttons, personal hygiene, scratching, and using electronics. Concerns about user safety and privacy regulations might influence the low number of works related to personal hygiene and scratching. In contrast, using electronics or phones is becoming more convenient with smart home applications and hands-free technology. In conclusion, the chosen task for the presentation aligns well with the literature. More focus should be placed on assisting people with personal hygiene tasks. The variety of tasks performed demonstrates that the participants were not influenced by the tasks shown in the video.

\subsection{Functionality} 
\label{subsec:DiscussionFunctionality}
\subsubsection{Would the prototype eye-tracking controller work in the controlled environment of the lab?} 
Initial testing showed that 70\% of object selections would result in selection success when the dwell time trigger is set to 500\,ms. Although dwell times can be shortened to enhance selection success, as shown in other literature \cite{Holmqvist2017}, this could lead to an unintentional object selection, as described in Section \ref{subsec:BackgroundSharedControl}. An appropriate threshold must be explored with the prototype system.

Accuracy issues are one of the main reasons why gaze was tracked outside of an AoI. Therefore, the influences on AoI misses and dwell time were investigated in Section \ref{subsec:ResultsRealWorldStudy}. Four dependencies were found: \begin{itemize}
    \item Object size: As shown in Figure \ref{fig:figure7}, object size correlates significantly with dwell time. Small objects, like keys, are part of everyday activities. Excluding such objects from interaction would lead to major drawbacks in participation in daily life. The available workspace is limited by the length of the robotic arm. This also limits the relative size of the object in the camera scene. With this limit, the resulting accuracy error can be calculated and checked with the desired interactive object size to estimate if this would cause problems. In this case, the accuracy of the eye-tracker is $\Delta_{deg}=0.6$\textdegree. With an effective workspace of $d=89.18$\,cm for the used robot arm \cite{KinovaGen3}, the maximum accuracy error is 9.34\,mm ($\Delta = d*tan(\Delta_{deg})$). Since all selected objects were larger than 9.34\,mm, it is assumed that the prototype would work.
    \item Geometry: The AoI of the fork could lead to a failed selection due to its width. Currently, there are two approaches in planning for estimating object outline. Object recognition can be used to create bounding boxes, solving the issue due to the new geometry. Overlapping of bounding boxes may occur more often, leading to unintentional object selection and context misinterpretation. Object segmentation can lead to boundaries which fit the object's outline. In this case, the geometry of the object would be traced with high accuracy, solving the bounding box issue, leading back to the initial width issue.
    \item Gaze location towards the object: In the eating task example, participants tend to move their heads to the left and right instead of up and down (see Figure \ref{fig:figure8}). The resulting higher accuracy error was due to the glasses periphery. Considering physical disabilities of potential users, an immobility of the head cannot be excluded. 
    \item User intention: Participants were instructed to focus their gaze on the selected object, but not on which region in detail. In the case of the bottle, the task was to fill the glass. The bottle needs to be opened first, which explains why the bottle cap was fixated, indicating the hand-eye coordination and intention behind the task, as discussed in Section \ref{subsec:BackgroundSharedControl}.
\end{itemize}

\subsubsection{What technical difficulties arise from the design?}

Ambiguous object selection can occur due to object overlap and was found in the case of the fork (see Figure \ref{fig:figure6}). In addition to the accuracy challenges, robust object selection may be the most difficult challenge in implementing the prototype. Other researchers have addressed this issue using 3D gaze tracking to minimize localization errors in approaches involving eye-tracking glasses \cite{Cao2025,Yang2021}. As discussed in Section \ref{subsec:BackgroundSharedControl}, user intent may lead to more robust selection methods by involving multiple objects to estimate task intent. These objects may vary in geometry and size, which reduces the impact of incorrect fixations. Regarding current approaches using object detection, overlap in bounding boxes of closely arranged objects may lead to larger issues than those found in the area of interest (AoI) evaluation. The AoIs were fitted to the outlines of the objects, similar to object segmentation. However, bounding boxes have larger areas, which leads to ambiguity when they overlap.

Section \ref{subsec:BackgroundCharacteristics} presents technical challenges, level of autonomy, and user preferences as factors influencing user acceptance. Within these categories, we found indications from the survey and hands-on study of technical difficulties.

After implementing the suggested control method, the tasks will be selectable as planned. However, as Hagengruber et al. \cite{Hagengruber2025} point out, accomplishing subsequent tasks is necessary to realize benefits for everyday usage. With this in mind, objects could indicate multiple tasks. Solutions integrating machine learning methods to identify gaze patterns, as presented by Yang et al., Jain et al., and Belardinelli, can reduce misinterpretation \cite{Yang2021, Jain2019, Belardinelli2024}.

In the WoZ design, the robot was presented with a high level of autonomy. As discussed in Section \ref{subsec:BackgroundCharacteristics}, it is questionable whether participants will accept the system as intended. In the hands-on study, participants vocalized their safety concerns, especially in the group with no prior robot experience. Arbitration methods can improve perception. This is especially important, since robot perception changes with regular use, and the level of autonomy can be reduced until the user is confident enough to operate the robot with a higher level of autonomy. Additionally, settings should be integrated that allow users to adjust the robot's velocity to their preferred level.

\subsection{Differences between Online Study and Hands-On User Study}
\label{subsec:DiscussionPerception}
\subsubsection{Were the participants satisfied with the presented system?}
In both studies, the participants were mostly satisfied with the system. We found in both studies suggestions for improvement. The design can be refined to meet the participants' expectations of the system. The framework must have the following features to realize the system:

\begin{itemize}
    \item Gaze detection on objects: Due to the outcomes in Section \ref{subsec:DiscussionFunctionality}, several approaches are followed to detect objects in the scene camera of the eye-tracking glasses. The prototype will estimate unintentional selection and context misinterpretation. 
    \item Task availability: The survey showed that the desired tasks were in line with the participants' expectations. Eating was the most frequently mentioned task. However, handling a fork is a safety risk when the robot manipulates the fork autonomously, and has to be handled appropriately. 
    \item Task selection: The expected prototype will implement objects associated with multiple tasks. Testing has shown that if the dwell time is set correctly and the algorithm is designed to wait for multiple objects to be selected, the task should be robustly selectable. Dwell times of 500 ms and less were found to be appropriate. As mentioned in Section \ref{subsec:BackgroundSharedControl}, feedback strategies need to be evaluated and integrated into the control to ensure the user. Strategies have to be evaluated if user intent methods assist in solving issues selecting between multiple tasks.
    \item Trajectory, grasping, and collision avoidance: The robot has to perform the task by itself. In the current version, the robot's trajectory was programmed, and the objects were placed at the same locations for each test. In a real-world scenario, this would not meet the user's expectations. Solutions were found, as presented in Section \ref{subsec:BackgroundSharedControl}), and depend on the approach chosen for object gaze detection.     
    \item Adaptation: Due to the ambiguous results of the robot's velocity perception, settings will be included in the prototype so that users can adapt the system to their liking, such as robot velocity. 
\end{itemize}

In future work, the experimenter's control has to be automatized so that the robot can autonomously perform tasks selected by the user.

\subsubsection{Does perception vary between the hands-on study and the online survey?}
The unstructured, in-person interviews with participants in the hands-on study and the open-ended questions in the online survey provided more diverse feedback on this shared control design. The results are consistent with, and extend, current literature in this area. It adds information for the design of gaze-controlled assistive robotic arms. Perceptions of the system do not vary, but highlight different aspects of the design.

Involving the stakeholder group yielded insights about daily tasks and their daily life with their disability. The prior-experience group revealed potential drawbacks in the technical design process, such as concerns about safety, kinematic selection, and trajectory planning. Although the non-prior-experience group did not have in-depth feedback on certain aspects of life with a disability or technology, they helped build a larger pool of participants to measure gaze behavior. In addition, their perception of the robot was more curious and, at the same time, hesitant than that of the prior-experience group due to the novelty of robot interaction. Since robotic arms are rarely available to PSMD, it is reasonable to assume that, in a real-world test, PSMD would respond in an initial robot interaction similarly to the non-prior experience group.

\section{Limitations} 
\label{sec:Limitations}
\textbf{Online study:} The qualitative analysis of the online survey provided limited insight into the participants' everyday lives, meaning the data cannot be generalized to a larger population. To counter this issue, we enriched and compared our information with suitable literature. For future studies, however, we have decided to conduct interviews with participants to minimize potential hurdles due to a lack of assistive devices or display problems on end devices. Q4 was tailored to evaluate whether hurdles exist. No indications were found. Additionally, no emails stating technical hurdles were received from medical professionals, family, or PSMD. The online survey might have benefited from additional questions. The safety concerns were not reflected in the survey responses. Including questions such as "Would you feel safe interacting with the robot?" would have provided a more comprehensive understanding of this topic. However, other studies show that potential users are more likely to accept the risks of using the system than reject it, becoming dependent on medical professionals or family members \cite{Baumeister2021}. This can also be seen here, as the group with robot experience is aware of the risks of human-robot interaction, but is not fearful of interacting with the robot. 
The survey did not ask about the occupation of the medical professionals. Therefore, opinions may differ depending on the occupational field (e.g., doctors, caregivers, and nurses). The online survey was designed for quantitative analysis. Due to the limited number of participants, a qualitative analysis was performed. Therefore, not all answers were discussed in detail, since they are not representative (e.g., Q5 rel./fr.: time spent on care). These results can be accessed in the supplementary materials. As discussed in Section \ref{subsec:MethodOnlineSurvey}, participants did not interact directly with the robot, which might affect perception.

\textbf{Specific Wizard of Oz design:} The Wizard of Oz system does not vary in intent interpretation due to its hard-coded trajectories. Variable behavior was excluded from this study, since comparing outcomes between participants would not be applicable. However, varying behavior in shared control changes the perception of the system, as shown in \cite{Aronson2018}. We presented a design driven by eye-tracking, in which objects are selected without the need for a GUI. This approach is uncommon, since prototype systems face challenges such as interpreting the desired task, estimating errors due to head movements, and lacking feedback modalities \cite{FischerJanzen2024}. If differences in control design exist in relation to the WoZ design, the results may not be transferable to other controls. 

\textbf{Generalized approach:} Since this approach was not tailored to specific disabilities, it may lack the details necessary to adapt to the needs of individual users. Future work can use this generalization to more easily identify these requirements.

\textbf{Transfer to other eye trackers:} Only the Tobii Pro Glasses 3 eye tracker was used, which is more accurate than other eye-tracking glasses on the market. However, the effect on the bounding boxes in object detection can vary greatly depending on the model used. Due to the scope of this paper, the test was not reproduced with other eye trackers.

\section{Conclusion}  
The goal of this work was to obtain feedback on an eye-tracking-driven shared control for assistive robotic arms in order to identify community expectations, desired tasks, and technical challenges. While previous studies offer valuable findings, they vary between robot types and involve alternative stakeholder groups. As a result, the available insights are generally only limited transferable and not specific to the control for an assistive robotic arm examined here. As suggested by other researchers, a Wizard of Oz design was created prior to prototyping. The factors necessary for successfully implementing shared control were identified and presented. 

Based on this WoZ design, two studies were conducted. An online survey designed for PSMD, their families and friends, and medical professionals. This survey provided feedback on the community’s expectations for the system and the tasks the robot should be able to perform. A hands-on study with participants with and without robotics experience provided insight into the technical requirements and feedback differences between the video presentation available for the online survey and the real-world application. Gaze data was recorded to measure dwell times and influence dependencies related to the eye-tracking controls. 

The duality of both studies provided insight into the usability and functionality of the system. Task prioritization was contextualized with current literature and approaches that researchers presented to realize shared controls. Dependencies were uncovered for the design of eye-tracking-driven robot controls, such as the influence of object size, geometry, gaze location, and user intent on accuracy. The impact on object detection or online task intent recognition based approaches were assessed and discussed. Consequently, improvements and open challenges related to the initial idea were presented. 

\begin{acks}
Both studies were approved by the ethics commission of the Offenburg University of Applied Sciences. 
The study designs were checked to comply to the ACM Publications Policy on Research Involving Human Participants and Subjects.
The Tobii Pro Glasses 3 were funded by the Deutsche Forschungsgemeinschaft (DFG, German Research Foundation) – project number INST898/33-1.
Further, we would like to thank Atikkhan Faridkhan Nilgar and Karl Tschurtschenthaler for the great discussions in statistics and eye-tracking evaluation related to this topic. We would like to thank Katrin-Misel Ponomarjova for proofreading the translations of the survey. We want to thank Kim Zähringer for her assistance with the robot and video making.
\end{acks}

\clearpage

\appendix
 \section{Overview of Questions asked in the Online Survey}
\label{app:appendix1}

In the following an Overview of the questions asked to the participants in the online Survey are presented. They represent three versions of the survey which would be handed to the individual participant group PSMD, family and friends and medical professionals. All related answers can be found in the supplementary materials.

\setlength{\LTcapwidth}{\textwidth}
\begin{longtable}{p{1.5cm}p{3.6cm}p{3.6cm}p{3.6cm}}
\caption{This table presents all questions asked in the online survey. Questions spanning over all three columns indicate questions that were asked to each participant.}
\label{tab:tab_OverviewQuestions} \\
    \toprule
        Indicator  & \multicolumn{3}{c}{Questions asked to group} \\
                            & PSMD & Relatives and friends & Medical professionals\\
\hline
        Q1                    & \multicolumn{3}{c}{What age group are you in?} \\
        Q2                    & \multicolumn{3}{c}{What gender are you?} \\
        Q3                    & \multicolumn{3}{c}{To which of the groups below do you feel you belong? } \\
        Q4                    &Do you have assistance from another person in completing this questionnaire? &Did the person you care for participate in this survey?& Please estimate the percentage of your patients or clients who have movement difficulties in their arms?\\
        Q5                    & If so, what is the reason you need support? &How much time do you spend on care each day (in hours)?&Let's assume that an assistance system could take over all daily tasks. Which tasks would you most like to hand over in order to facilitate care? Name 1-3 favorites. \\
        Q6                    &How often does your impairment affect your daily life?&Do you already use assistive devices to help you provide care?& -\\
        Q7                    &In terms of your physical disability, has your condition improved, worsened, or stayed the same in the last 6 months?&If yes, which assistive devices do you use?& -\\
        Q8                    &Select the body movements that you can control without restrictions&-& -\\
       Q9                    &Do you already use a robotic gripper arm, for example on an electric wheelchair? & \multicolumn{2}{p{7.2cm}}{Does the person(s) you care for already use a robotic gripping arm, e.g. on an electric wheelchair?} \\
        Q10                    &If so, how do you control it? & \multicolumn{2}{p{7.2cm}}{If so, how does the person control this robotic arm?} \\
        Q11                    &Are you satisfied with your current system?&\multicolumn{2}{p{7.2cm}}{Are you satisfied with the system in terms of startup and function?}\\
        Q12                    &If no, indicate why you are not satisfied with the system.&\multicolumn{2}{p{7.2cm}}{If no, why are you not satisfied with the system?}\\
        Q13                    &In relation to your disability, would it be possible for you to use the system presented in the video every day?&\multicolumn{2}{p{7.2cm}}{In which tasks would you or the person you care for most like to be supported by this system in daily care? Please name your 1-3 favorites.}\\
        Q14                    &If no, what changes would you suggest to make the system easier for you to use?&\multicolumn{2}{p{7.2cm}}{Would it be possible for the person you care for to use the system presented in the video every day?}\\
        Q15                    &In which tasks would you most like to be supported by this system? Please name your 1-3 favorites.&\multicolumn{2}{p{7.2cm}}{If no, what changes would you suggest to make the system easier to use for you and the person you care for?}\\
        Q16                    & \multicolumn{3}{c}{Do you have any other comments on the questions asked or the system?} \\
        Q17                    & \multicolumn{3}{p{10.8cm}}{Would you like to have had more detailed information on how to use the system to answer the questions asked here?} \\
        Q18                    & \multicolumn{3}{c}{If yes, what information would you have liked to have?} \\
          \bottomrule
\end{longtable}

 \section{Participant statements of hands-on study}
\label{app:appendix2}
In the following the statements of participants of the hands-on study are listed. All statements were translated by the authors to english. Participants without any statement to the system were not listed below.
\setlength{\LTcapwidth}{\textwidth}
\begin{longtable}{p{0.5cm}p{1.7cm}p{10.5cm}}
  \caption{Statements of participants. Prior-experience group: Exp., Non-Prior-experience group: Non-Exp.}
  \label{tab:tab3} \\
    \toprule
    P & Exp./ Non-Exp. & Statement\\
    \midrule
    1      & Non-Exp. & By noting that the robot does not go over the table, P1 was reassured. “apprehension, causious” system if not had stated, Glass's task: fear of overspilling and dropping the cup in the real task.  \\
    3         & Non-Exp. & Food task: Lifting a fork so high that you only had to grab was cool. Light switch was hit centered, that was cool. The handle of the cup was held towards the person. That was very practical if you wanted to grab it.\\
    4       & Non-Exp. & Fork task: Test person was confused because the fork pointed to a fixed point and did not go in the direction of the mouth. General: Confused as to whether the robot cancels the task when looking at another object.\\
    5    & Non-Exp.  &  Movements too slow for everyday use. However, it was satisfactory for an initial test.\\
    6         & Non-Exp. & Food task: Serving food was interesting. Glass task: The glass slipped a little when I put it down. There was no straw in the glass, so it wasn't clear why the robot handed the glass flat. \\
    7      & Non-Exp. & Open gripper is uncomfortable when it points at you. Velocity was good.  \\
    10      & Non-Exp. & General: By saying that the robot does not reach over the table, the action was not scary. \\
    11    & Non-Exp.  & General: Responds quickly, which is good. Intuitive operation. Once you know where the robot stops it is less intimidating/scary. \\
    13      & Exp. &  Fork task: First close the gripper a little before gripping the fork completely, less risk of crushing the hand. General: Emergency stop is poorly positioned.\\
    14    & Exp.  & Kinematics moves spongily, drinking task: Kinematics has touched the bottle.\\
    15         & Exp. & When the fork is dropped, a loud unpleasant noise is heard from the drop. \\
    16      & Exp. &  P16 asked whether tasks are aborted if you accidentally look at other objects. Waiting time is appropriate until the robot moves. It does what you want (intuitive control). Kinematics vibrate during movement, which is negative.\\
    17    & Exp.  & Robot is shaky, jerks. Selection process: Fear of triggering the wrong action when looking at the wrong object.\\
    18         & Exp. & Movement was a little too fast to the light switch \\
    22      & Exp. & With pick and place, it was assumed that either the block or the platform could be gripped. Accordingly, P22 was confused. \\
    23    & Exp.  & Robot trembles, trajectory not optimal, bottle badly twisted while handled.\\
    24        & Exp. & Fork was recognized well, even though you looked briefly at the bottle during the time you had to aim. The pause between aiming and starting the pick and place task was too long.\\
  \bottomrule
\end{longtable}

\bibliographystyle{ACM-Reference-Format}
\bibliography{sample-base}

\end{document}